\documentclass[journal]{IEEEtran}

\usepackage{times}

\usepackage{soul}
\usepackage[hidelinks]{hyperref}
\usepackage[utf8]{inputenc}
\usepackage[small]{caption}
\usepackage{graphicx}
\usepackage{amsmath}
\usepackage{booktabs}
\usepackage{mathrsfs}
\usepackage{amsfonts,amssymb}
\usepackage{multirow}
\usepackage{color}
\usepackage{cite}
\usepackage{pifont}
\usepackage{array}
\usepackage[misc]{ifsym} 

\usepackage{makecell}
\usepackage{textcomp}
\usepackage{makecell}

\begin{document}
\title{Lung Nodule Segmentation and Uncertain Region Prediction with an Uncertainty-Aware Attention Mechanism}
\author{
  Han~Yang, Qiuli~Wang, Yue~Zhang, Zhulin An, Liu~Chen, Xiaohong~Zhang, and S.~Kevin Zhou
  \thanks{Corresponding author: Qiuli~Wang and S.~Kevin Zhou. }
  \thanks{Q.~Wang, Y.~Zhang are with 
  School of Biomedical Engineering, Division of Life Sciences and Medicine,
  University of Science and Technology of China, Hefei, Anhui, 230026, China
  and
  Center for Medical Imaging, Robotics, Analytic Computing \& Learning (MIRACLE),
  Suzhou Institute for Advanced Research, University of Science and Technology of China, Suzhou, Jiangsu, 215123, China.
  E-mail: $\{$wangqiuli, yue\_zhang$\}$@ustc.edu.cn
  }
  \thanks{H.~Yang is with the Institute of Computing Technology, Chinese Academy of Sciences, Beijing 100190, China and also with University of Chinese Academy of Sciences, Beijing 100049, China . E-mail: yanghan22s@ict.ac.cn.}
    \thanks{Z.~An is with the Institute of Computing Technology, Chinese Academy of Sciences, Beijing 100190, China. E-mail: anzhulin@ict.ac.cn.}
  \thanks{C.~Liu is with Department of Radiology, The First Affiliated Hospital of Army Medical University, Chongqing 400032, China. E-mail: liuchen@aifmri.com.}
  \thanks{X.~Zhang is with School of Big Data \& Software Engineering, Chongqing University, Chongqing 401331, China. E-mail: xhongz@cqu.edu.cn.}
  \thanks{S.K. Zhou is with School of Biomedical Engineering, Division of Life Sciences and Medicine,
  University of Science and Technology of China, Hefei, Anhui, 230026, China
  and
  Center for Medical Imaging, Robotics, Analytic Computing \& Learning (MIRACLE),
  Suzhou Institute for Advanced Research, University of Science and Technology of China, Suzhou, Jiangsu, 215123, China
  and also with the Key Laboratory of Intelligent Information Processing of Chinese Academy of Sciences (CAS), Institute of Computing Technology, CAS, Beijing 100190, China. E-mail: skevinzhou@ustc.edu.cn.}
  \thanks{H.~Yang and Q.~Wang contributed equally to this paper. }
}

\maketitle
\begin{abstract}
  Radiologists possess diverse training and clinical experiences, leading to variations in the segmentation annotations of lung nodules and resulting in segmentation uncertainty. Conventional methods typically select a single annotation as the learning target or attempt to learn a latent space comprising multiple annotations. However, these approaches fail to leverage the valuable information inherent in the consensus and disagreements among the multiple annotations.
  In this paper, we propose an Uncertainty-Aware Attention Mechanism (UAAM) that utilizes consensus and disagreements among multiple annotations to facilitate better segmentation.
  To this end, we introduce the Multi-Confidence Mask (MCM), which combines a Low-Confidence (LC) Mask and a High-Confidence (HC) Mask. The LC mask indicates regions with low segmentation confidence, where radiologists may have different segmentation choices. Following UAAM, we further design an Uncertainty-Guide Multi-Confidence Segmentation Network (UGMCS-Net), which contains three modules: a \emph{Feature Extracting Module} that captures a general feature of a lung nodule, an \emph{Uncertainty-Aware Module} that produces three features for the annotations' union, intersection, and annotation set, and an \emph{Intersection-Union Constraining Module} that uses distances between the three features to balance the predictions of final segmentation and MCM.
  To comprehensively demonstrate the performance of our method, we propose a Complex Nodule Validation on LIDC-IDRI, which tests UGMCS-Net's segmentation performance on lung nodules that are difficult to segment using common methods. Experimental results demonstrate that our method can significantly improve the segmentation performance on nodules that are difficult to segment using conventional methods.
\end{abstract}

\IEEEpeerreviewmaketitle

\section{Introduction}
\IEEEPARstart{L}{ung} nodule segmentation is crucial in Computer-Aided Diagnosis (CAD) systems for lung cancer \cite{wang2017central}, providing critical information such as nodule sizes, shapes, and other important medical features.
However, for the general train-and-test paradigm of deep learning methods, each nodule image data has only one annotation mask delineated by one radiologist \cite{shariaty2022texture, agnes2022efficient,oktay2018attention,ronneberger2015u,zhang2021deeprecs}. Thus, the network can only provide a single prediction of nodule regions each time.

However, in clinical practice, different radiologists may provide various segmentation annotations for a lung nodule due to their various training and clinical experience \cite{hu2019supervised, kohl2019hierarchical, xiaojiang2021}. As a result, the traditional methods based on single annotation cannot reflect the diversity of clinical experiences and limit the application of deep learning methods.

A straightforward solution to the problem of varying annotations among radiologists is to incorporate multiple annotations for each lung nodule image.
That leads to another issue: multiple annotations inevitably bring uncertainty and conflicts, as radiologists may annotate the same regions differently.
To overcome this problem, Kohl \textit{et al.} proposed a probabilistic U-Net in 2018, which utilized a conditional variational auto-encoder to encode multiple segmentation variants into a low-dimensional latent space \cite{kohl2018probabilistic, kohl2019hierarchical}. 
By sampling from this space, the network could affect the corresponding segmentation map.
Based on this study, Hu \textit{et al.} proposed to combine the ground-truth uncertainty with a probabilistic U-Net, which could improve predictive uncertainty estimates, sample accuracy, and sample diversity \cite{hu2019supervised}.
These methods relied on the latent space and random samples in this space. Consequently, these methods could only provide uncertain regions by multiple predictions.

\begin{figure}[htbp]
  \centerline{\includegraphics[width=80mm]{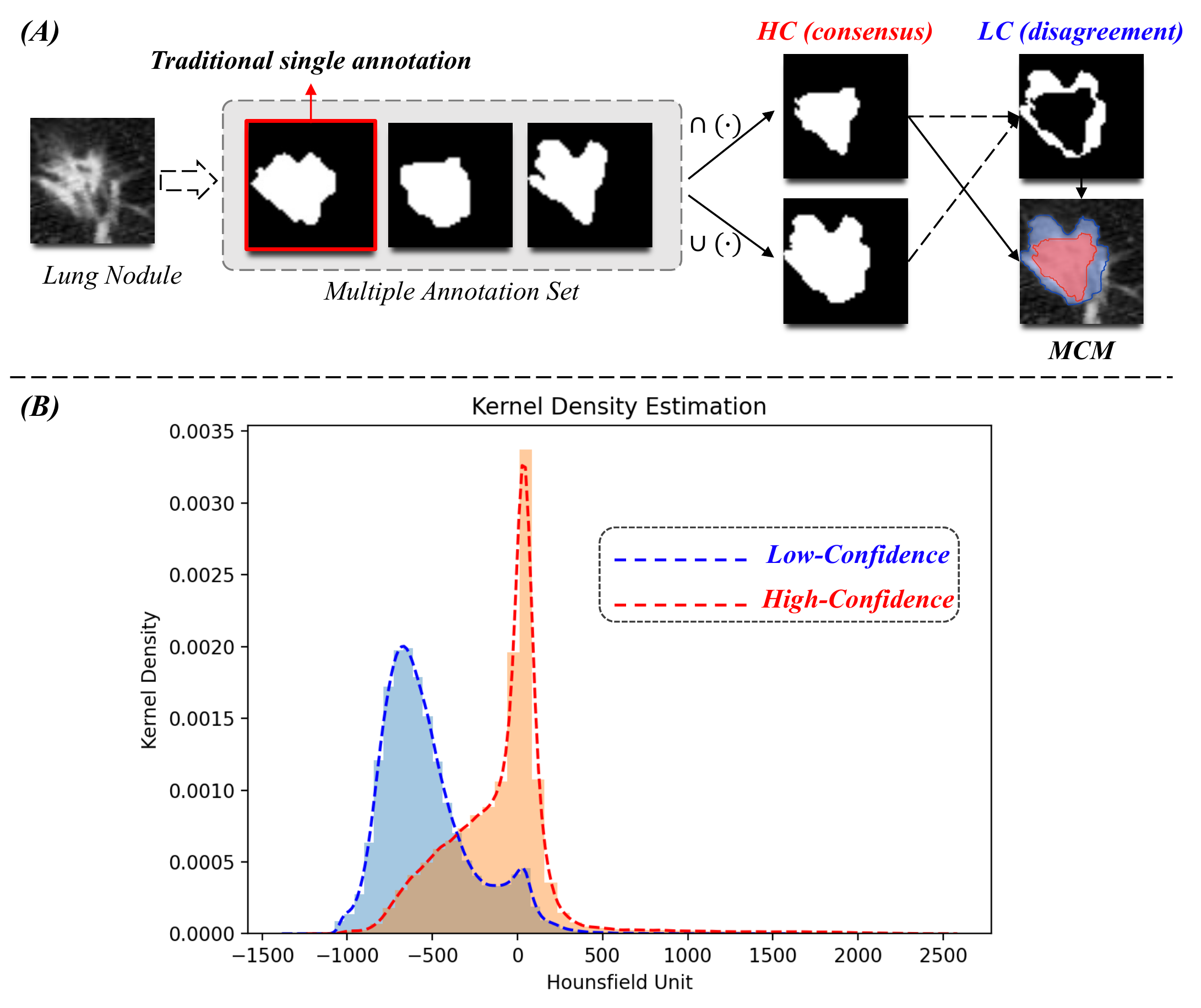}}
  \vspace{-0.0cm}
  \caption{(A). Uncertainty caused by multiple annotations. HC is High-Confidence Mask, which is the intersection of the annotation set. LC is Low-Confidence Mask, which is the difference between the annotation set's union and intersection. Multi-Confidence Mask is the combination of LC and HC.
  (B). Hounsfield Unit Kernel Estimations in HC and LC of LIDC-IDRI.}
  \vspace{-0.0cm}
  \label{MCM}
\end{figure}

In this paper, we present an argument that \textbf{the uncertainty between multiple annotations follows a particular pattern}. To demonstrate this phenomenon, we introduce the Multi-Confidence Mask (MCM), which combines a High-Confidence (HC) Mask and a Low-Confidence (LC) Mask, as illustrated in Figure~\ref{MCM}. A. 
\textbf{The intersection mask is equal to the HC mask, representing the intersection of all annotations. The union mask is the union of all annotations. The LC mask is the difference between intersection mask and union mask.}
When calculating the Hounsfield Unit (HU) Kernel Estimations of HC and LC on the LIDC-IDRI dataset \cite{armato2011lung}, as shown in Figure~\ref{MCM}. B, we can observe a noticeable distinction in the HU distribution between the LC and HC masks. Specifically, the LC regions have lower HU values than the HC regions.
From a pixel distribution perspective, a lower HU value indicates a lower density of the corresponding region. In terms of CT image features, the LC regions predominantly consist of boundary-related features like nodule edges, spiculation, and ground-glass features, whereas the HC region is primarily distributed within the core of the nodule. As a result, we put forth the hypothesis that \textbf{the areas leading to discrepancies among radiologists are primarily linked to low-density tissue and boundary-related features.}

\begin{figure}[htbp]
  \centerline{\includegraphics[width=80mm]{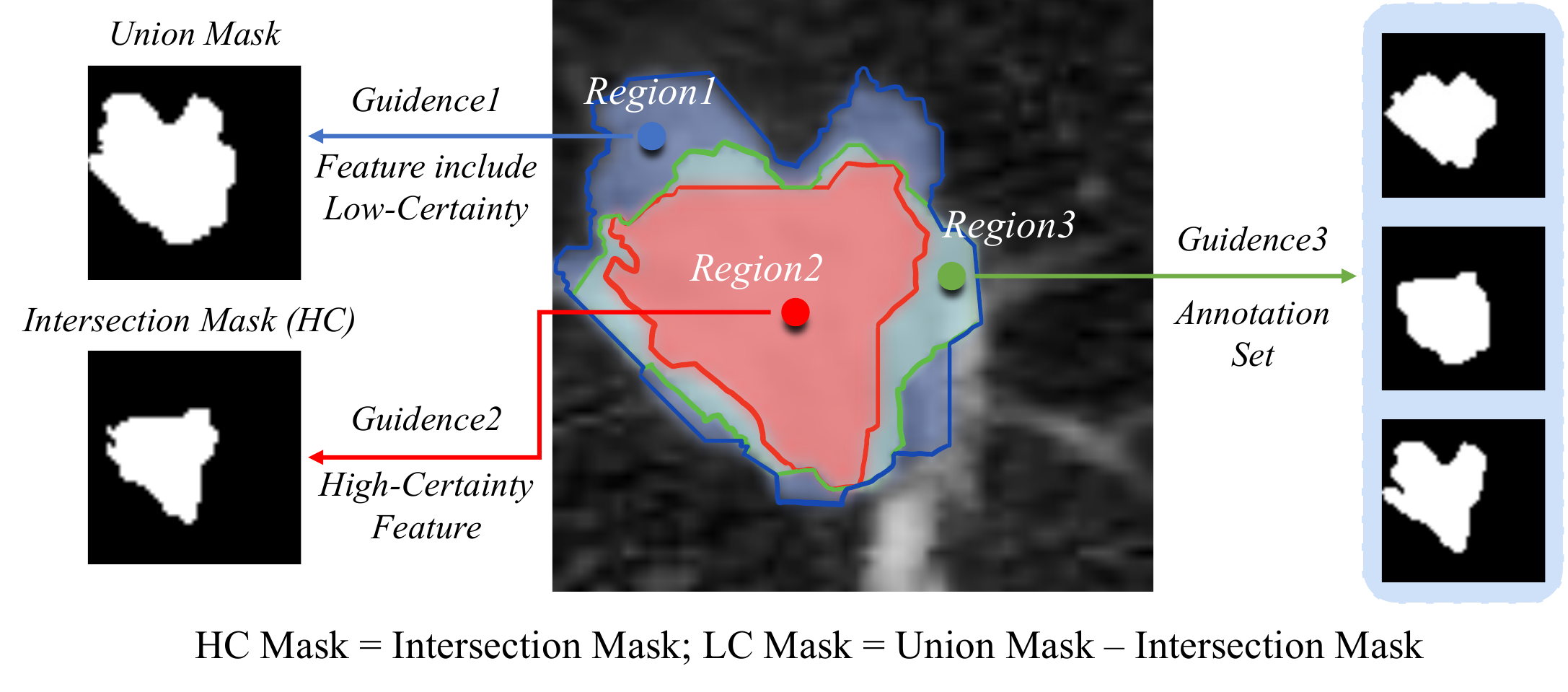}}
  \vspace{-0.0cm}
  \caption{Overview of \emph{Uncertainty-Aware Attention Mechanism}. The difference between union and intersection masks is the Low-Confidence Mask, which guides the learning of low-certainty features.
  The intersection mask is the High-Confidence Mask, which guides the learning of high-certainty features.
  Meanwhile, the annotation set guides the learning of a plausible segmentation, which is a balance between all annotations.}
  \vspace{-0.0cm}
  \label{UAM}
\end{figure}

Unlike other methods, we propose utilizing the MCM and annotation set as the learning guidance of features with different segmentation certainties, contributing to better segmentation performance. We call this training \emph{Uncertainty-Aware Attention Mechanism} (UAAM) hereafter, which is illustrated in Figure~\ref{UAM}. Following this mechanism, we further design an Uncertainty-Guide Multi-Confidence Segmentation Network (UGMCS-Net) for lung nodule segmentation.

The UGMCS-Net contains three modules: a U-Net-based \emph{Feature Extracting Module}, an \emph{Uncertainty-Aware Module}, and an \emph{Intersection-Union Constraining Module}.
Firstly, the \emph{Feature Extracting Module} extracts general feature maps from the input CT image.
Secondly, the \emph{Uncertainty-Aware Module} transfers the general feature maps into three independent feature maps $R_{LC}$, $R_{HC}$ and $R_{Uni}$ with the guidance of annotations' union, intersection, and the annotation set. 
$R_{LC}$ and $R_{HC}$ are used to predict the union $\cup(X)$ and intersection $\cap(X)$, and the results are combined as MCM. $R_{Uni}$ is used to predict $X_{Uni}$. 
Thirdly, the \emph{Intersection-Union Constraining Module} captures preferred features with Feature-Aware Attention Blocks from $R_{LC}$, $R_{HC}$ and $R_{Uni}$, then constrains the final prediction $X_{S}$ with feature distances, ensuring the segmentation results are constrained in a moderate and reasonable way. To better utilize multiple annotations, we also introduce a Multiple Annotation Fusion Loss to optimize the $X_{Uni}$ and $X_{S}$, which calculates the average BCE loss between prediction and all annotations.

The proposed method has two distinct advantages:
(1) Instead of learning from a latent space, the proposed method has specific learning targets, enabling it to provide a stable prediction of uncertain nodule regions.
(2) The method optimizes the prediction with all annotations to ensure that the final prediction balances different conditions, making the most of the available information.

We have reported a preliminary version of this work in a previous publication \cite{yang2022uncertainty}. 
The new contributions of this paper can be summarized as follows:  
\begin{itemize}
  \item 
  \textbf{A novel mechanism called the \emph{Uncertainty-Aware Attention Mechanism} (UAAM)}: UAAM maximizes the utility of multiple annotations and employs the Multi-Confidence Mask (MCM) to guide the learning of low and high-confidence features. In contrast to conventional VAE-based methods, UAAM can deliver stable predictions for uncertain regions and single lung nodule segmentations together.
  \item \textbf{An upgraded Uncertainty-Guide Multi-Confidence Segmentation Network (UGMCS-Net)}: Based on the mechanism, we update the UGS-Net to the UGMCS-Net, which contains a \emph{Feature Extracting Module}, an \emph{Uncertainty-Aware Module}, and an \emph{Intersection-Union Constraining Module}. To make the most of multiple annotations, we introduce a Multiple Annotation Fusion Loss, comparing the segmentation with all possible annotations. The proposed modules are plug-and-play, which can be applied to other segmentation networks in different situations. 
  \item \textbf{A comprehensive validation}: We propose a Complex-Nodule Validation, which tests UGMCS-Net's segmentation performance on the lung nodules that are difficult to segment by U-Net. Experiments demonstrate that for nodules whose DSC score is lower than 60$\%$ on U-Net, the DSC score of our network can be increased by 11.03$\%$, and the IoU score of our network can be improved by 11.88$\%$. We also provide sufficient ablation studies for different modules, backbones, and model settings.
\end{itemize}

\section{Related Work}
\label{relatedwork}
\subsection{Lung Nodule Segmentation}
Lung nodule segmentation is critical in lung nodule computer-aided detection (CAD) systems. Its primary goal is to accurately delineate the boundary of a target nodule to provide details such as its diameter, size, and semantic features \cite{wu2010stratified,gonccalves2016hessian,pezzano2021cole,xie2019automated}.
The main challenge of this task is that lung nodules have various shapes, sizes, and delicate features. In the early years, researchers provided numerous methods for lung nodule segmentation, such as morphology-based methods and region-growing-based methods \cite{diciotti2011automated,dehmeshki2008segmentation}. Recently, deep learning has become the most popular method in this area.

In 2017, Wang \emph{et al.} proposed a multi-view convolutional network for lung nodule segmentation.
The proposed network captured a diverse set of nodule-sensitive features from axial, coronal, and sagittal views in CT images simultaneously. 
These features were analyzed with a multi-branch CNN network, achieving an average DSC similarity coefficient (DSC) of 77.67\% \cite{wang2017multi}. 
Also, in 2017, Wang \emph{et al.} proposed a central-focused convolutional neural network with central-pooling layers to analyze nodules in 2D and 3D thoroughly \cite{wang2017central}. In 2020, Cao \emph{et al.} designed a dual-branch residual network with intensity-pooling layers, which enhanced the learning of intensity information and improved the DSC to 82.74\% \cite{cao2020dual}. In 2021, Pezzano \emph{et al.} introduced a CNN network that could learn the context of the nodules by producing two masks representing all the background and secondary-important elements in the CT, so that the network can better discriminate nodule features \cite{pezzano2021cole}.
Later in 2022, Shariaty \emph{et al.} further proposed texture feature extraction and feature selection algorithms to improve the segmentation, achieving a DSC of 84.75\% \cite{shariaty2022texture}.

Based on the observations from the aforementioned studies, it is evident that existing methods have predominantly prioritized achieving more precise segmentation, while overlooking the fact that different radiologists may hold various opinions on how to segment the same lung nodule. In this study, we argue that the disagreements among annotations also hold diagnostic value. Therefore, our method aims to generate a segmentation that effectively balances all annotations by learning from annotation sets and identifies regions with varying segmentation certainties.

\subsection{Uncertainty in Lung Nodule Segmentation}
Many medical image vision problems suffer from ambiguities. In clinical situations, it may not be clear from a CT scan alone which particular region is cancer tissue \cite{kohl2018probabilistic,9558816}. As a result, even experienced doctors and radiologists may provide different segmentations for the same tissues or tumor. 

In 2018, Kohl \emph{et al.} proposed to model this task as the learning of distribution over diverse but plausible segmentations for lung nodules. Based on U-Net \cite{ronneberger2015u}, they introduced a probabilistic U-Net, which was a combination of a U-Net and a conditional VAE that could produce an unlimited number of plausible segmentations. Later in 2019, Kohl \emph{et al.} further proposed a hierarchical probabilistic U-Net, which used a hierarchical latent space decomposition to formulate the sampling and reconstruction of segmentations with high fidelity \cite{kohl2019hierarchical}. Also in 2019, Hu \emph{et al.} analyzed two types of uncertainty: aleatoric and epistemic \cite{hu2019supervised}. They exploited multiple annotations' variability as a source of `ground truth' aleatoric uncertainty, combined this uncertainty with probabilistic U-Net, and tried to quantitatively analyze the segmentation uncertainty. In 2021, Long \emph{et al.} extended the concept in \cite{hu2019supervised} to the V-Net and 3D lung nodule CT images. As an ideal dataset, which contained multiple annotations for over 1000 lung nodules, all these studies \cite{kohl2018probabilistic,kohl2019hierarchical,hu2019supervised,xiaojiang2021} analyzed LIDC-IDRI.

Different from the networks based on VAE, our work pays more attention to the reason that causes the various annotations, which manifest as segmentation divergence. We introduce an alternative method that specifically targets uncertainty regions, enabling us to make stable predictions for both uncertain nodule regions and overall lung nodule segmentation. This approach allows us to gain insights into the underlying causes of segmentation discrepancies and produce more reliable results in uncertain lung nodule regions.

\section{Methods}
\label{method}
\subsection{Uncertainty-Guided Multi-Confidence Segmentation Network}
\begin{figure*}[htbp]
  \centerline{\includegraphics[width=150mm]{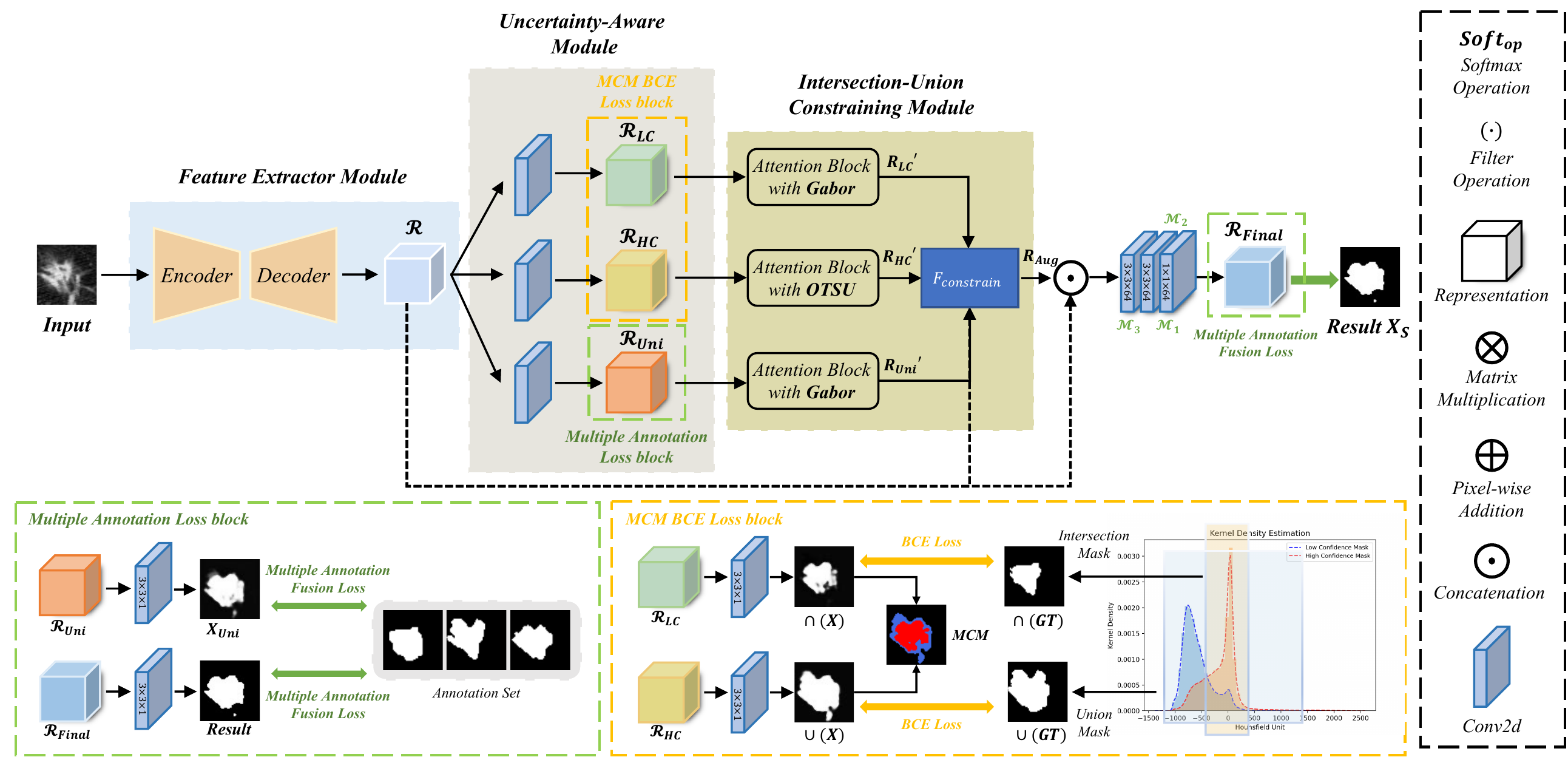}}
  \vspace{-0.0cm}
  \caption{
  Overview of Uncertainty-Guided Multi-Confidence Segmentation Network. This network contains three modules:
  (1) \emph{Feature Extracting Module}, (2) \emph{Uncertainty-Aware Module}, and (3) \emph{Intersection-Union Constraining Module}.}
  \vspace{-0.0cm}
  \label{arch}
\end{figure*}
  
In Figure~\ref{arch}, we present the architecture of the Uncertainty-Guided Multi-Confidence Segmentation Network (UGMCS-Net). 
The network takes the lung nodule CT image as inputs, and produces two outputs: a predicted Multi-Confidence Mask (MCM) and a final segmentation $X_{S}$. 
The MCM combines predicted union $\cup(X)$ and intersection $\cap(X)$.
The learning targets of the network are the annotation set $GT$, as well as their Union Mask $\cup(GT)$ and Intersection Mask $\cap(GT)$. The input images and their corresponding masks have dimensions of $50\times50$ pixels, obtained through cropping from the LIDC-IDRI dataset with official annotations.
Before being fed into the network, the input images and masks are resized to dimensions of $3\times64\times64$ pixels.

The UGMCS-Net contains three modules:
(1) \emph{Feature Extracting Module}, (2) \emph{Uncertainty-Aware Module}, and (3) \emph{Intersection-Union Constraining Module}.
The \emph{Feature Extracting Module} can use any segmentation network based on U-Net structure to initially obtain a feature map $R$ with a shape of $32\times64\times64$.
This paper uses Attention U-Net \cite{oktay2018attention} with five down-sampling and up-sampling layers. Each up-sampling layer is composed of two convolutional layers and an attention block. 
The \emph{Uncertainty-Aware Module} analyzes $R$ and generates $R_{LC}$, $R_{HC}$, and $R_{Uni}$. These feature maps are then fed into MCM BCE Loss Block and Multiple Annotation Loss Block, producing the initial $\cup(X)$, $\cap(X)$, and a plausible segmentation $X_{Uni}$. The union $\cup(X)$ and intersection $\cap(X)$ are computed to obtain the MCM.
The \emph{Intersection-Union Constraining Module} learns the different characteristics of $R_{LC}$, $R_{HC}$, and $R_{Uni}$, and fuse these three features into $R_{Fianl}$. This module then provides a more reasonable final segmentation $X_{S}$ by analyzing $R_{Fianl}$.

\subsection{Uncertainty-Aware Module}
\label{UAM_method}
The \emph{Uncertainty-Aware Module} (UAM) is introduced to make full and reasonable use of all annotation information by learning $\cup(GT)$, $\cap(GT)$, and $GT$. This module has two tasks: (1) capturing different features from Low-Confidence (LC) regions, High-Confidence (HC) regions, and all annotations; (2) producing initial predictions of the Multi-Confidence Mask (MCM) and a general segmentation.

As shown in Figure~\ref{arch}, UAM adopts a three-branch CNN network as its backbone.
It takes $R$ ($32\times64\times64$) as input and extracts $R_{LC}$, $R_{HC}$, and $R_{Uni}$ using three different convolution layers with a kernel size of $1\times1$. The $R_{LC}$, $R_{HC}$, and $R_{Uni}$ have a same size of $32\times64\times64$.
The MCM BCE Loss Block receives $R_{LC}$ and $R_{HC}$, generating $\cup(X)$ and $\cap(X)$ with three different convolution layers with a kernel size of $3\times3$.
The BCE loss is calculated between $\{\cup(X), \cup(GT)\}$ and $\{\cap(X), \cap(GT)\}$. The $\cup(X)$ and $\cap(X)$ are combined as $MCM'$ through the normalization operation $Normal(\cup(X)+\cap(GT))$, which reflects the degree of uncertainty in different regions. 
Unlike our previous work \cite{yang2022uncertainty}, the branch for $R_{Uni}$ is optimized by the Multiple Annotation Loss Block, which will be discussed later. 
Additionally, feature maps $R_{LC}$, $R_{HC}$, and $R_{Uni}$, which have the same shape of $1\times64\times64$, will be fed into the next module for further analysis.

\subsection{Intersection-Union Constraining Module}
\label{SCM}
As mentioned above, $\cup(GT)$ and $\cap(GT)$ are the learning targets of UAM. Specifically, $\cup(GT)$ indicates all regions that might be nodule tissues, and $\cap(GT)$ indicates nodule regions with the highest confidence. To achieve a balance between extreme situations, we further develop a new module called the \emph{Intersection-Union Constraining Module} (IUCM). This module is designed to capture the features of all three learning targets, and produce a more reasonable segmentation prediction that can strike a balance between extreme situations.

\begin{figure}[htbp]
  \centerline{\includegraphics[width=70mm]{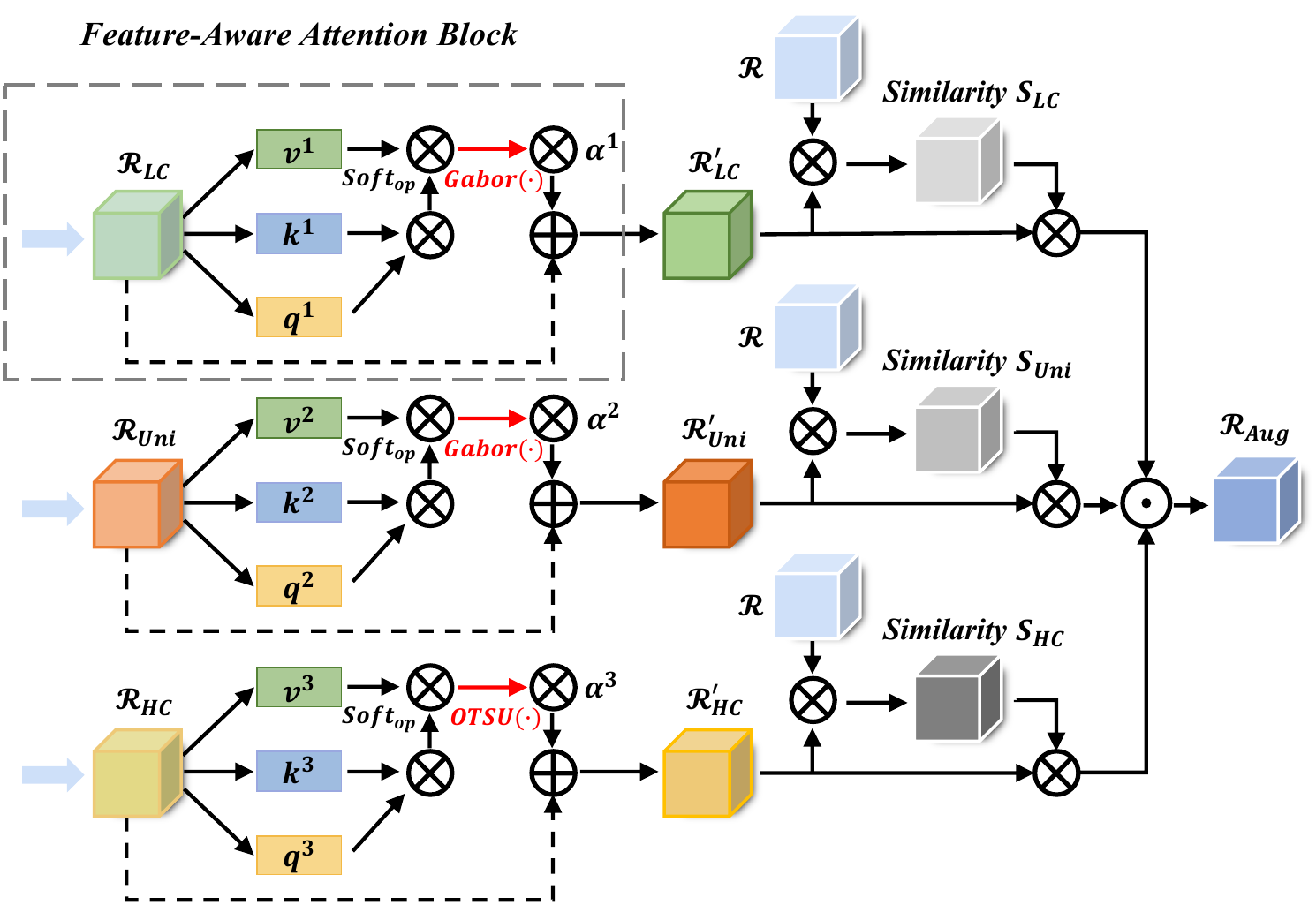}}
  \vspace{-0.0cm}
  \caption{
    Intersection-Union Constraining Module. It contains three Feature-Aware Attention Blocks, and each block has specific feature learning preferences.
    }
  \vspace{-0.0cm}
  \label{UAM_F}
\end{figure}

As shown in Figure~\ref{UAM_F}, IUCM takes $R_{LC}$, $R_{HC}$, and $R_{Uni}$ as its inputs, and generals corresponding $R'_{LC}$, $R'_{HC}$, and $R'_{Uni}$ with Feature-Aware Attention Block (FAAB). 
$R'_{LC}$, $R'_{HC}$, and $R'_{Uni}$ have the same size of $32\times32\times32$.
FAAB is built based on a self-attention block \cite{vaswani2017attention} and a feature-aware filter.
These attention blocks process $R_{LC}$, $R_{HC}$, and $R_{Uni}$ with different feature-aware filters, which enable the network to formulate different learning preferences for different learning objectives and obtain more image features that are helpful for segmentation.
More specifically, assuming the input is $R_z$, the process of FAAB can be summarized as:

\begin{equation}
  \begin{aligned}
    R'_z=R_z\oplus \Gamma(A(R_z))
  \end{aligned}
\end{equation}
where $z\in\{Uni,LC,HC\}$, $A$ indicates the self-attention architecture. $\Gamma$ is a feature-aware filter, in this study, the $\Gamma$ for $R_{Uni}$ and $R_{LC}$ is Gabor \cite{luan2018gabor}, the $\Gamma$ for $R_{HC}$ is Otsu \cite{2007A}. 
The $\Gamma(A(R_z))$ is pixel-wise added with $R_z$ so that the network can keep more information from the input.

Through the observation of the dataset and Hounsfield Unit Kernel Estimations, we can see that $R_{HC}$ is mainly solid nodules with higher density, while $R_{LC}$ includes more low-density tissues (such as burrs), which are mainly distributed in the edges of nodules. 
The Otsu is sensitive to density characteristics, which can help the network identify high-density tissue more accurately. So, we apply Otsu to extract $R'_{HC}$ from $R_{HC}$. 
Meanwhile, Gabor is sensitive to image edges and can provide good direction selection and scale selection features, so that it can capture local structural features in multiple directions in the local area of the image. 
Therefore, we choose Gabor to extract $R'_{LC}$ and $R'_{Uni}$ from $R_{LC}$ and $R_{Uni}$. 
The ablation study about filter choices will be provided in Section~\ref{experiments}.

After getting $R'_{LC}$, $R'_{HC}$, and $R'_{Uni}$, the IUCM gets $S_z=d\{R'_z,R\}$, $d$ is the operation for calculating cosine similarity. 
The output of IUCM is $R_{Aug}=Concat(S_{Uni}\times R'_{Uni},S_{LC}\times R'_{LC},S_{HC}\times R'_{HC})$. $R_{Aug}$ will be concatenated with $R$ from the Feature Extracting Module, fed into a convolutional layer, and produce the final segmentation prediction $X_{S}$. The concatenation of $R$ and $R_{Aug}$ keeps more information from CT inputs.

\subsection{Loss Function}
\label{Loss}
In Figure~\ref{arch}, the UGMCS-Net contains two optimizations: MCM BCE Loss Block and Multiple Annotation Loss Block.

MCM BCE Loss Block calculates BCE loss between $\{\cup(X), \cup(GT)\}$ and $\{\cap(X), \cap(GT)\}$, which can be represented as:
\begin{equation}
  \begin{aligned}
    L_{MCM}=BCE(\cup(X), \cup(GT))+\\BCE(\cap(X), \cap(GT))
  \end{aligned}
\end{equation}

\begin{figure}[htbp]
  \centerline{\includegraphics[width=40mm]{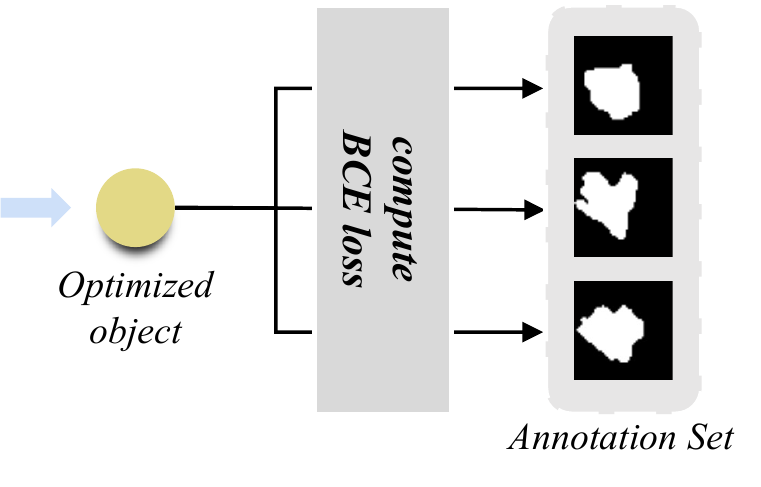}}
  \vspace{-0.0cm}
  \caption{
    Multiple Annotation Fusion Loss.
  }
  \vspace{-0.0cm}
  \label{maloss}
\end{figure}

We use the \emph{Multiple Annotation Fusion Loss} to optimize $X_{Uni}$ and $X_{S}$ in the Multiple Annotation Loss Block, denoted as $\Phi$. In our previous work, only one annotation from the set was selected for optimizing $X_{Uni}$ and $X_{S}$, which resulted in a loss of valuable information from other annotations. This study introduces the Multiple Annotation Fusion Loss, which compares the prediction with all possible annotations.
Firstly, $R_{Uni}$ and $R_{Fianl}$ produces $X_{Uni}$ and $X_{S}$. Secondly, as illustrated in Figure~\ref{maloss}, Multiple Annotation Fusion Loss functions calculate the BCE losses between the optimized objects ($X_{Uni}$ and $X_{S}$) with annotation set, and incorporate overage of these losses. This approach enables us to better utilize the information from all annotations in the set.
We have:
\begin{equation}
  \begin{aligned}
    \Phi_a=E\{\Sigma_{j}BCE (X_{Uni},GT_j) \} \\
    \Phi_b=E\{\Sigma_{j}BCE (X_{S},GT_j) \}
  \end{aligned}
\end{equation}
where $GT_j\in GT$. 

The loss fusion of the network can be defined as:
\begin{equation}
  \begin{aligned}
    L=\alpha_{1}L_{MCM}+\alpha_2\Phi_a+\alpha_3\Phi_b
  \end{aligned}
\label{eq_weight}
\end{equation}
where $\alpha_1$, $\alpha_2$, and $\alpha_3$ are pre-defined parameters. 
Empirically, $\alpha_1$ is set to 0.5, $\alpha_2$ to 0.5, and $\alpha_3$ to 1 in this paper.
Ablation study for weight selection will be displayed in Section~\ref{experiments}.

\section{Experiments}
\label{experiments}
\subsection{Dataset and Experimental Settings}
\label{settings}
In this study, we evaluate the proposed network using the LIDC-IDRI dataset \cite{armato2011lung}, which consists of 1018 study instances and over 2600 nodules.
For this study, we select 1860 lung nodules with a diameter of 3-30mm and multiple annotations. 
Each nodule data has at least two annotations so that we can obtain its CT image, multiple annotation set $GT$, their union $\cup(GT)$, and intersection $\cap(GT)$. 
The input images and their masks are both $50\times50$ pixels, which are cropped from the LIDC-IDRI dataset based on the official annotations. 
Each pixel reflects Hounsfield Unit of CT images.
We clip the intensity values of CT images to the range [-1000,1000] and normalize all intensity values to the range [0, 1] before training. The code for the data process is provided on the website \href{https://github.com/qiuliwang/LIDC-IDRI-Toolbox-python}{https://github.com/qiuliwang/LIDC-IDRI-Toolbox-python}.

We conduct our experiments on a server running Ubuntu 18.04 with a Tesla V100 GPU, using CUDA 11.2 with about 16G GPU memory. The network is implemented with PyTorch-v1.0.1 and Python3.7. We use five-fold validation to evaluate the network's effectiveness, ensuring robustness to data splits. We use Stochastic Gradient Descent with Warm Restarts (SGDR) as the optimizer, with an initial learning rate (LR) of 0.00001, a batch size of 32, a momentum of 0.9, and a weight decay rate of 0.0001. Each network is trained for 200 epochs, with the learning rate updated every 50 epochs.
The source code of UGMCS-Net, the original USG-Net, and all experimental settings will be uploaded to the \href{https://github.com/yanghan-yh/UGS-Net}{https://github.com/yanghan-yh/UGS-Net}.

\subsection{Performance of Lung Nodule Segmentation}
\label{performanceana}
\begin{table*}[h]\tiny
  \setlength{\tabcolsep}{5pt}
  \renewcommand\arraystretch{1.1}
  \vspace{-0.0cm}
  \caption{Performance comparison between our UGMCS-Net and nine networks based on the U-Net structure on the LIDC-IDRI dataset.
  UGS-Net represents a preliminary version of this work in a \cite{yang2022uncertainty}.
  All indicators are expressed in percentages.}
  \centering
  \begin{tabular}{c|ccc|ccc|ccc|ccc|ccc|ccc}		
  \hline
  \hline
  \multirow{2}*{\textbf{\emph{Method}}} & \multicolumn{3}{c}{\textbf{\emph{Fold1}}} & \multicolumn{3}{c}{\textbf{\emph{Fold2}}} & \multicolumn{3}{c}{\textbf{\emph{Fold3}}} & \multicolumn{3}{c}{\textbf{\emph{Fold4}}} & \multicolumn{3}{c}{\textbf{\emph{Fold5}}} & \multicolumn{3}{c}{\textbf{\emph{Average}}}\\
  \cline{2-19} 
  ~ & \multicolumn{1}{c}{\textbf{\emph{DSC}}} & {\textbf{\emph{IoU}}} & {\textbf{\emph{NSD}}} & \multicolumn{1}{c}{\textbf{\emph{DSC}}} & {\textbf{\emph{IoU}}} & {\textbf{\emph{NSD}}} & \multicolumn{1}{c}{\textbf{\emph{DSC}}} & {\textbf{\emph{IoU}}} & {\textbf{\emph{NSD}}} & \multicolumn{1}{c}{\textbf{\emph{DSC}}} & {\textbf{\emph{IoU}}} & {\textbf{\emph{NSD}}} & \multicolumn{1}{c}{\textbf{\emph{DSC}}} & {\textbf{\emph{IoU}}} & {\textbf{\emph{NSD}}} & \multicolumn{1}{c}{\textbf{\emph{DSC}}} & {\textbf{\emph{IoU}}} & {\textbf{\emph{NSD}}} \\
  \hline
  \makecell[l]{FCN \cite{DBLP:journals/corr/LongSD14}} & 83.92 & 73.51 & 92.63 & 84.80 & 74.75 & 93.64 & 85.37 & 75.35 & 94.81 & 84.54 & 74.44 & 94.21 & 85.74 & 75.78 & 94.80 & 84.87 $\pm$ 0.64 & 74.77 $\pm$ 0.78 & 94.02 $\pm$ 0.82 \\
  \makecell[l]{U-Net \cite{DBLP:journals/corr/RonnebergerFB15}}  & 84.98 & 75.18 & 93.12 & 86.34 & 76.89 & 93.67 & 87.26 & 78.21 & 95.50 &86.20 & 76.67 & 94.83 & 86.52 & 77.02& 95.17 & 86.26 $\pm$ 0.74 & 76.79 $\pm$ 0.97 & 94.46 $\pm$ 0.91 \\
  \makecell[l]{R2U-Net \cite{DBLP:journals/corr/abs-1802-06955}} & 83.33 & 72.92 & 92.10 &85.66 & 76.02 & 94.17 &86.44 & 77.01 & 95.23 & 84.83 & 74.86 & 93.38 & 85.61 & 75.66 & 95.26 & 85.17 $\pm$ 1.05 & 75.29 $\pm$ 1.37 & 94.03 $\pm$ 1.19 \\
  \makecell[l]{R2AU-Net \cite{R2AU-Net}} & 82.99 & 72.36 & 91.50 & 83.84 & 73.43 & 92.92 &86.13 & 76.50 & 94.58 & 85.10 & 75.24 &94.24 &85.41 & 75.53 & 95.03 & 84.69 $\pm$ 1.13 & 74.61 $\pm$ 1.50 & 93.65 $\pm$ 1.27 \\
  \makecell[l]{Attention U-Net \cite{DBLP:journals/corr/abs-1804-03999}} &85.99 & 76.44 & 93.98 & 86.11 & 76.58 & 93.68 & 87.76 & 78.89 & 96.07 & 86.50 & 77.13 & 95.42 & 86.99 & 77.61 & 95.56 & 86.67 $\pm$ 0.65 & 77.33 $\pm$ 0.88 & 94.94 $\pm$ 0.94 \\
  \makecell[l]{Nested U-Net \cite{DBLP:journals/corr/abs-1807-10165}} & 84.72 & 74.70 & 93.12 & 85.40 & 75.61 & 94.39 & 86.06 & 76.35 & 95.57 & 85.50 & 75.54 & 95.16 & 86.31 & 76.62 & 95.90 & 85.60 $\pm$ 0.55 & 75.76 $\pm$ 0.68 & 94.83 $\pm$ 0.99 \\
  \makecell[l]{Channel U-Net \cite{10.3389/fgene.2019.01110}} & 85.28 & 75.44 & 94.15 & 86.02 & 76.58 & 93.22 & 86.93 & 77.62 & 95.94 & 86.25 & 76.82 & 95.36 & 86.56 & 76.99 & 95.46 & 86.21 $\pm$ 0.56 & 76.69 $\pm$ 0.71 & 94.82 $\pm$ 1.00 \\
  \makecell[l]{nnU-Net \cite{nnUnet}} & 85.52 & 75.68 & - & 84.40 & 73.99 & - & 85.05 & 75.06 & - & 83.91 & 73.61 & - & 84.13 & 73.91 & - & 84.60 $\pm$ 0.60 & 74.45 $\pm$ 4.80 & - \\
  \makecell[l]{UGS-Net \cite{yang2022uncertainty}} & 86.47 & 77.07 & 93.58 & 87.21 & 78.17 & 95.26 & 88.39 & 79.86 & 96.20 & 86.56 & 77.17 & 95.56 & 87.19 & 77.94 & 95.78 & 87.16 $\pm$ 0.69 & 78.04 $\pm$ 1.00 & 95.28 $\pm$ 0.90 \\ 
  \hline
  \makecell[l]{UGMCS-Net} & 87.49 & 78.43 & 94.67 & 87.72 & 78.86 & 95.43 & 88.66 & 80.31 & 96.50 & 86.93 & 77.86 & 95.64 & 87.44 & 78.44 & 95.84 & \textbf{87.65 $\pm$ 0.57}\textcolor{green}{$\uparrow$}& \textbf{78.78 $\pm$ 0.83}\textcolor{green}{$\uparrow$} & \textbf{95.62 $\pm$ 0.59}\textcolor{green}{$\uparrow$} \\ 
  \hline
  \hline
  \end{tabular} 
  \label{tabel1}
  \vspace{-0.0cm}
\end{table*}

We compare the proposed UGMCS-Net with eight commonly used segmentation networks on traditional lung nodule segmentation. These networks include FCN \cite{DBLP:journals/corr/LongSD14}, U-Net \cite{DBLP:journals/corr/RonnebergerFB15}, R2U-Net \cite{DBLP:journals/corr/abs-1802-06955}, R2AU-Net \cite{R2AU-Net}, Attention U-Net \cite{DBLP:journals/corr/abs-1804-03999}, Nested U-Net \cite{DBLP:journals/corr/abs-1807-10165}, Channel U-Net \cite{10.3389/fgene.2019.01110}, nnU-Net \cite{nnUnet}, and UGS-Net \cite{yang2022uncertainty}.
We have trained UGS-Net and UGMCS-Net with multiple annotations, while the other networks have been trained using the traditional strategy of having only one correlated annotation $Label_1$ for each nodule, which is the first in the annotation set.
Three metrics are used to evaluate the predictive ability of the network to the lesion regions in this study: the average Dice Similarity Coefficient (DSC), Intersection over Union (IoU), and Normalized Surface Dice (NSD) \cite{ma2021abdomenct}. 

Table~\ref{tabel1} illustrates that UGMCS-Net achieves the highest scores in DSC, IoU, and NSD, with values of 87.65$\%$ ($\pm$0.56$\%$), 78.78$\%$ ($\pm$0.83$\%$), and 95.62$\%$ ($\pm$0.59$\%$), respectively.
Compared to U-Net, UGMCS-Net demonstrates improvements of 1.39$\%$, 1.99$\%$, and 1.16$\%$ across the three metrics, respectively.
Likewise, compared to Attention U-Net, UGMCS-Net achieves enhancements of 0.98$\%$, 1.45$\%$, and 0.68$\%$ for the respective metrics.
These outcomes highlight UGMCS-Net's superior segmentation performance, particularly underscored by the substantial increase in the NSD score, indicating its strong boundary feature segmentation capability.
Furthermore, UGMCS-Net exhibits considerable advancements across all indicators when compared to UGS-Net, with a rise of 0.49$\%$ in DSC, 0.74$\%$ in IoU, and 0.34$\%$ in NSD scores.
Additionally, the variances of the three metrics obtained from the five-fold cross-validation of UGMCS-Net are consistently smaller than those of UGS-Net, suggesting enhanced network stability through the integration of Multiple Annotation Fusion Loss and constraining operations.
nnU-Net is a popular network for the segmentation task. However, it only achieved 84.60$\%$ in DSC and 74.45 $\%$ in IoU. This is because the training of nnU-Net requires a large dataset. However, we only have 1860 nodule images in this task.

\begin{figure}[t]
  \centerline{\includegraphics[width=75mm]{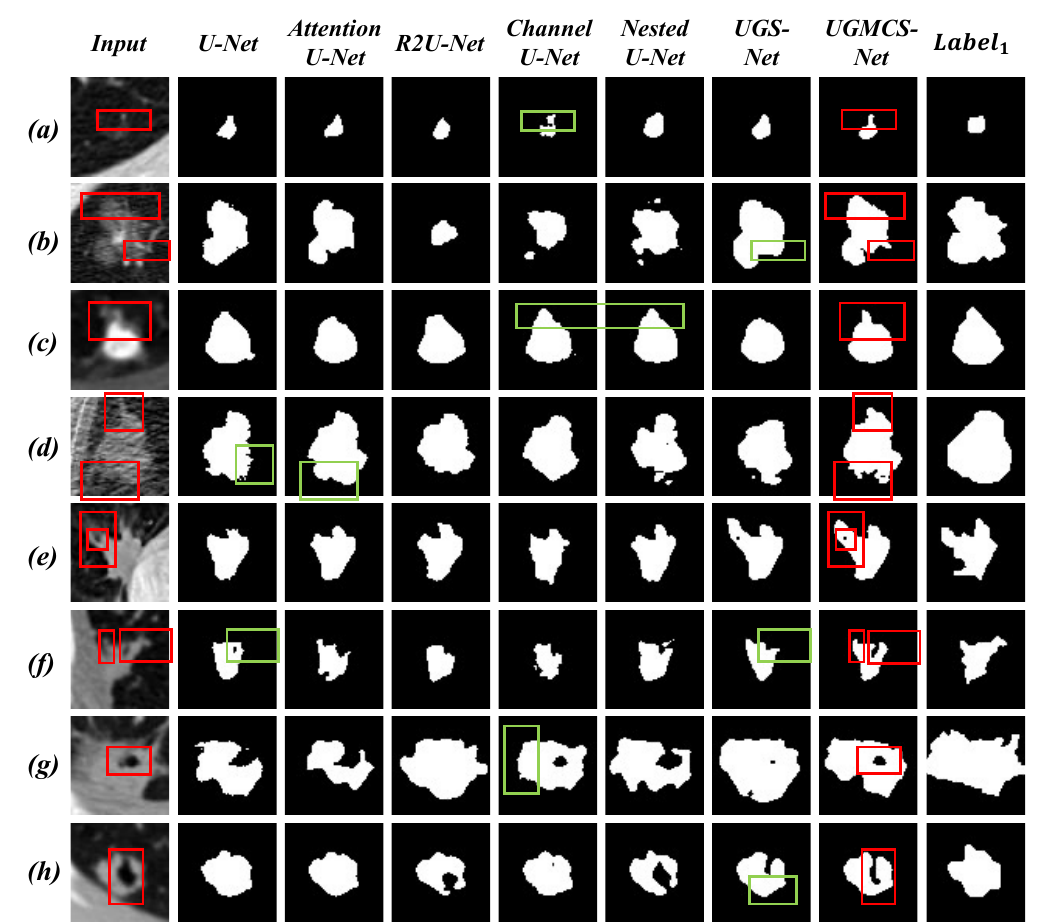}}
  \vspace{-0.0cm}
  \caption{Segmentation results of U-Net, Attention U-Net, R2U-Net, Channel U-Net, Nested U-Net, UGS-Net, and UGMCS-Net.
  The red boxes corresponding to the \textbf{\emph{Input}} column indicate the features that should be noted or the error-prone locations of the nodules during segmentation.
  The red boxes in the \textbf{\emph{UGMCS-Net}} column indicate the segmentation detail of UGMCS-Net at these locations.
  The green boxes indicate the inadequacies in the suboptimal segmentation result.}
  \vspace{-0.0cm}
  \label{segComparison}
\end{figure}

Figure~\ref{segComparison} shows partial segmentation results of the aforementioned methods. The red boxes in the \textbf{\emph{Input}} column indicate areas of interest or error-prone segmentation locations for the nodules. Nodules (a)-(c) contain many low-density regions, nodules (d)-(f) have irregular shapes, such as spiculation signs, at their boundaries, and nodules (g)-(h) have cavities. The segmentation of these areas by UGMCS-Net is significantly more consistent with the actual shape of the lesions than other methods.

Figure~\ref{segComparison} and Table~\ref{tabel1} demonstrate that: (1) Learning from the annotation set, as well as its union and intersection, provides richer visual information for the segmentation task. (2) Learning from LC areas improves the network's ability to recognize low-density regions.

\begin{figure*}[h]
  \centerline{\includegraphics[width=160mm]{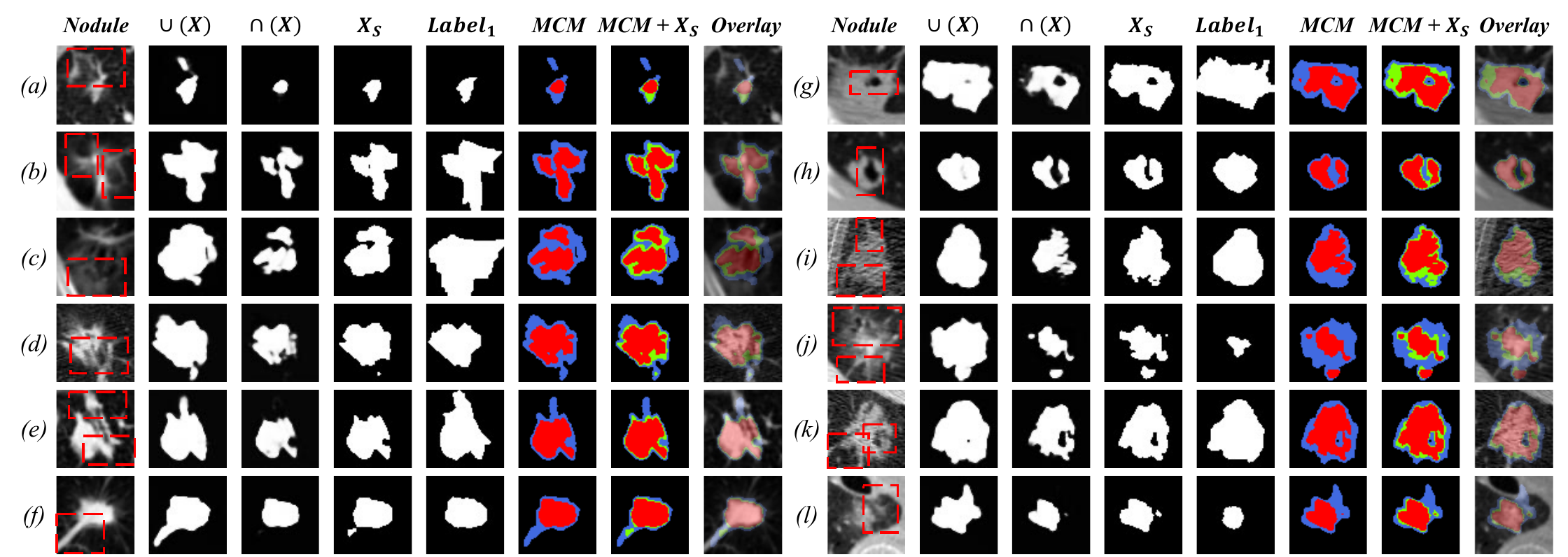}}
  \vspace{-0.0cm}
  \caption{The predicted intersection $\cap(X)$, predicted union $\cup(X)$, final segmentation $X_S$, and MCM are generated by the UGMCS-Net.
  Colors in MCM are used for better visualization, red for $\cap(X)$ and blue for $\cup(X)$.
  In addition, final segmentation is represented in the MCM and marked with green to facilitate comparison.
  Red boxes indicate areas or features of nodules that are not easily distinguishable.}
  \vspace{-0.0cm}
  \label{Experi_MCM}
\end{figure*}

\begin{table}[h]
  \setlength{\tabcolsep}{4pt}
  \renewcommand\arraystretch{1.2}
  \vspace{-0.0cm}
  \caption{Statistical analysis of U-Net, Attention U-Net, and UGMCS-Net.}
    \centering
  \begin{tabular}{c|cc|cc}		
  \hline
  \hline
  \multirow{3}*{\textbf{\emph{Method}}} & \multicolumn{4}{c}{\textbf{\emph{UGMCS-Net}}}\\
  \cline{2-5} 
    &\multicolumn{2}{c}{\textbf{\emph{Dice}}} & \multicolumn{2}{c}{\textbf{\emph{IoU}}}\\
  \cline{2-5} 
    &p-value & t-value & p-value & t-value\\
  \hline
  \makecell[l]{U-Net} & 2.80\texttimes$10^{-7}$& -5.15 & 1.02\texttimes$10^{-7}$ & -5.33\\
  \makecell[l]{Attention U-Net} & 5.53\texttimes$10^{-3}$ & -2.78 &  4.49\texttimes$10^{-3}$ & -2.84\\
  \hline
  \hline
  \end{tabular} 
  \label{tabel1_p}
  \vspace{-0.0cm}
\end{table}

It is worth noting that in Figure~\ref{segComparison} (g)-(h), there are no cavities in the annotation, but the segmentation result obtained by UGMCS-Net reflects the cavity feature. We choose to keep these features for two reasons:
(1) The common nodule tissues have a higher density than lung parenchyma. But cavities in lung nodules have extremely low density or are even empty. We can consider them as parts or not parts of nodules. 
(2) The segmentation maps should provide more information about nodule characters. The cavities are important characters, so keeping these cavities is a better choice.
The cavities in predictions are easy to remove using methods like cv2.findContours if needed.

In addition, we perform a statistical analysis of U-Net, Attention U-Net, and UGMCS-Net.
As shown in Table\ref{tabel1_p}, the p-value of UGMCS-Net for U-Net and Attention U-Net in Dice and IoU indexes is far less than 0.05, and the t-value is larger, indicating that UGMCS-Net is significantly better than U-Net and Attention U-Net in performance.

The experiments above are based on the first annotation $Label_1$ in the annotation set.
To eliminate the effect of mask choice, we also provide the performances of U-Net, Attention U-Net, and UGMCS-Net on $Label_2$, which is the second in the annotation set. 
The experimental results, which are listed in Table~\ref{tabel1_label2}, demonstrate that the UGMCS-Net keeps its superior performance on $Label_2$. It means the proposed method can keep its stable performance over different mask choices. 
In the conventional training approach, each nodule is assigned a single mask and fails to provide sufficient information for nodules characterized by complex structures. In our network, each nodule is associated with 2-4 masks. By integrating the Multiple Annotation Fusion Loss, we infuse more comprehensive information into the learning process. It is particularly beneficial for segmenting nodules with intricate structures and low-density textures. As a result, the Multiple Annotation Fusion Loss significantly enhances the performance, especially for nodules characterized by complex structures.

\begin{table}[t]
  \setlength{\tabcolsep}{4pt}
  \renewcommand\arraystretch{1.2}
  \vspace{-0.0cm}
  \caption{Performance comparison between U-Net, Attention U-Net, and UGMCS-Net on $Label_1$ and $Label_2$, which are presented as subscript 1 and 2.}
    \centering
  \begin{tabular}{c|cc|cc|cc}		
  \hline
  \hline
  \multirow{2}*{\textbf{\emph{Method}}} & \multicolumn{6}{c}{\textbf{\emph{Average $\%$}}}\\
  \cline{2-7} 
    &$DSC_1$&$DSC_2$ & $IoU_1$& $IoU_2$& $NSD_1$ & $NSD_2$\\
  \hline
  \makecell[l]{U-Net} &86.26& 86.46 &76.79& 77.10 &94.46& 94.92\\
  \makecell[l]{Attention U-Net} &86.67& 87.15 &77.33& 78.09 &94.94& 95.32\\
  \makecell[l]{UGMCS-Net} &87.65& 87.90 &78.78& 79.19 &95.62& 95.52\\
  \hline
  \hline
  \end{tabular} 
  \label{tabel1_label2}
  \vspace{-0.0cm}
\end{table}

\subsection{Uncertain Region Prediction}
Besides being able to segment nodules, UGMCS-Net can also predict regions more likely to be nodule tissues and regions with lower likelihood. Figure~\ref{Experi_MCM} illustrates the prediction outcomes $\cap(X)$ and $\cup(X)$, the final segmentation result $X_{S}$, and the generated MCM'. In MCM' and MCM+UGS-Net, red denotes the high-confidence region, blue signifies the low-confidence region, and green corresponds to the final segmentation $X_{S}$. In the ideal scenario, the $X_{S}$ should effectively strike a balance between the high-confidence and low-confidence regions.

According to Figure~\ref{Experi_MCM}, our final segmentation outcome lies between the two extreme scenarios of high and low confidence regions. These intermediate results demonstrate that (1) Our prediction is cognizant of all potential annotations, and (2) the prediction is constrained between $\cap(X)$ and $\cup(X)$. 

\begin{figure}[h]
  \centerline{\includegraphics[width=80mm]{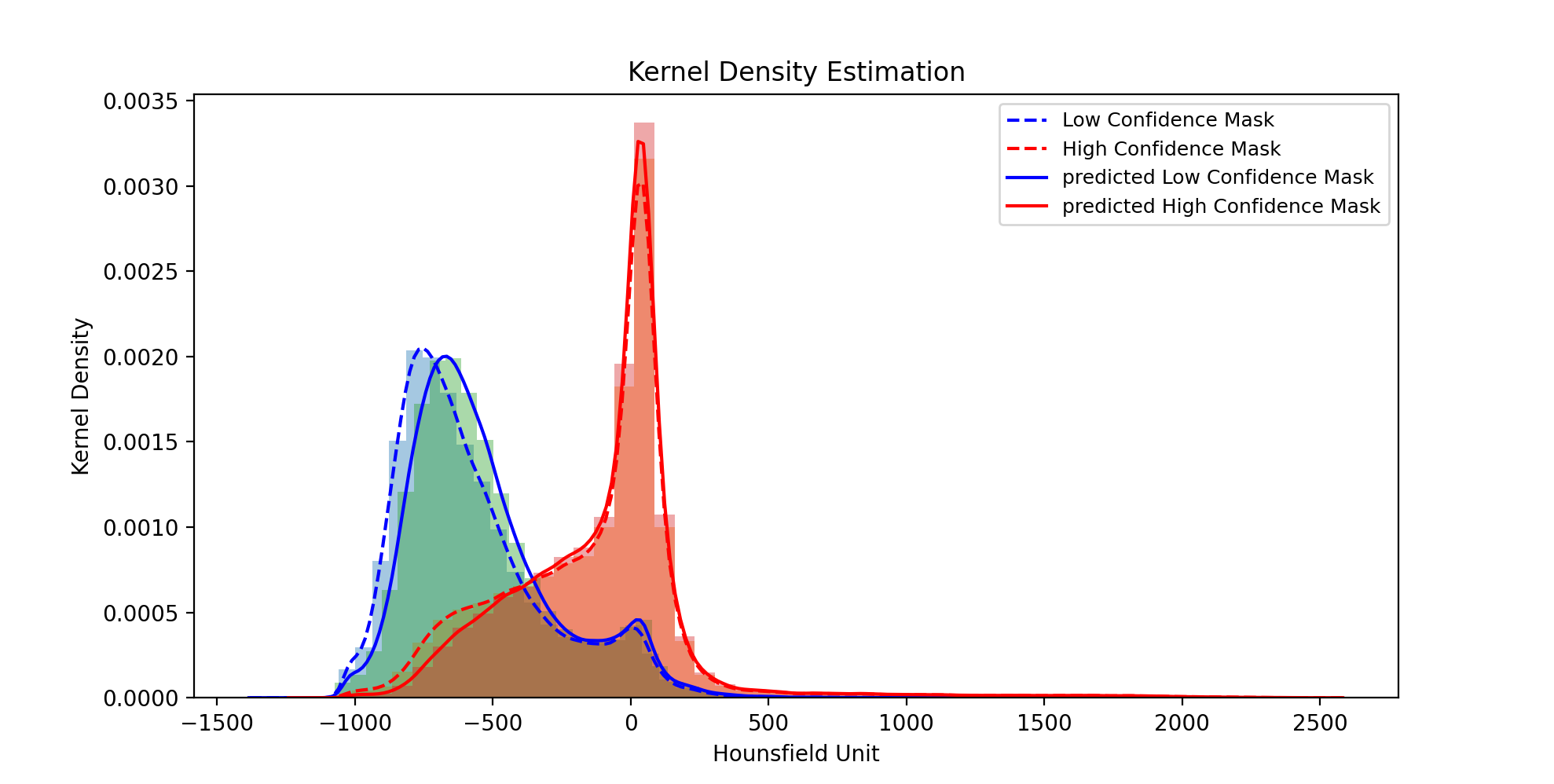}}
  \vspace{-0.0cm}
  \caption{Comparison of HU value kernel density estimation of real HC, LC, and predicted HC and LC.}
  \vspace{-0.0cm}
  \label{MCM_HU}
\end{figure}

Specifically, the use of MCM in lung nodule feature segmentation provides several advantages that can better display the semantic features of nodules:

(1) MCM can better highlight the significant cavity features of nodules (Figure \ref{Experi_MCM}.\emph{(g)(h)}). The nodule cavity features are more evident on the prediction mask with UGMCS-Net compared with other methods. This is due to UGMCS-Net's ability to capture the density difference of nodule tissue under the guidance of the intersection mask $\cap(GT)$ and union mask $\cup(GT)$, which helps to retain more features of the nodule cavity.

(2) MCM can better segment spiculation signs, which is an important feature for diagnosing benign and malignant lung nodules (Figure \ref{Experi_MCM}.\emph{(a)-(f)}). Spiculation is a stellate distortion caused by the intrusion of nodules into surrounding tissue and is typically low-density and distributed around the nodule edges. This feature is difficult to segment with traditional deep learning methods. UGMCS-Net can focus more on nodule boundary characteristics under the guidance of union mask $\cup(GT)$, resulting in better segmentation performance for spiculation distributed at the boundary of a nodule.

(3) MCM can better segment the low-density tissue of nodules (Figure \ref{Experi_MCM}.\emph{(i)-(l)}), which is commonly found in ground-glass nodules and is the main area that causes differences in expert labeling. UGMCS-Net can identify low-density tissue to the maximum extent through the study of the annotation set $GT$ and union mask $\cup(GT)$, which is very helpful for the diagnosis of ground-glass nodules.

Due to the small size of the LC mask, traditional quantitative evaluation metrics such as DSC and IoU are insufficient for measuring the prediction quality of MCM. To address this issue, we compare the HU distribution of the predicted HC and LC masks with the actual HC and LC masks. 
We assume that UGMCS-Net can reasonably predict the degree of uncertainty in different regions. Therefore, the HU value distribution of predicted HC and LC masks should be similar to the actual distribution. Figure \ref{MCM_HU} shows that the predicted curve is in good agreement with the actual curve, indicating that the level of regional uncertainty we predicted is statistically reliable.

\subsection{Ablation Study}
\subsubsection*{Ablation Study for Modules}
\label{ASModules}
We perform five-fold validation in each section if we do not point out practicality.
To better use the multiple annotations' information and enhance the relationship among $R_{LC}$, $R_{HC}$ and $R_{final}$, we update the UGMCS-Net with Multiple Annotation Fusion Loss and Intersection-Union Constraining Module.
To further demonstrate the contribution of these two modules, we constructed UGMCS-$\Phi_a$, UGMCS-$\Phi_b$, UGMCS-$\Phi_a$+$\Phi_b$, and UGMCS-IUCM for ablation experiments based on UGMCS-Net.
In UGMCS-$\Phi_a$, the Multiple Annotation Fusion Loss is only applied to the $\cup(X)$;
In UGMCS-$\Phi_b$, the Multiple Annotation Fusion Loss is only applied to the final output of the network $X_S$;
In UGMCS-$\Phi_a$+$\Phi_b$, the Multiple Annotation Fusion Loss is applied to both the $\cup(X)$ and $X_S$;
In UGMCS-IUCM, we use the Intersection-Union Constraining Module in the USG-Net, but no Multiple Annotation Fusion Loss.

\begin{table}[h]
  \renewcommand\arraystretch{1.2}
  \vspace{-0.0cm}
  \caption{Ablation Study for Modules. All indicators are expressed in percentages.}
  \centering
  \begin{tabular}{ccccccc}		
  \hline
  \hline
  \makebox[0.0003\textwidth][c]{\textbf{\emph{V1}}} & \makebox[0.0003\textwidth][c]{\textbf{\emph{$\Phi_{a}$}}} & \makebox[0.0003\textwidth][c]{\textbf{\emph{$\Phi_b$}}} & \makebox[0.01\textwidth][c]{\textbf{\emph{IUCM}}} & \makebox[0.03\textwidth][c]{\textbf{\emph{$DSC$}}} & \makebox[0.03\textwidth][c]{\textbf{\emph{$IoU$}}} & \makebox[0.03\textwidth][c]{\textbf{\emph{$NSD$}}} \\
  \hline
  - &  &  &  & 86.67 $\pm$ 0.65 & 77.33 $\pm$ 0.88 & 94.94 $\pm$ 0.94 \\
  \checkmark &  &  &  & 87.16 $\pm$ 0.69 & 78.04 $\pm$ 1.00 & 95.28 $\pm$ 0.90 \\
  \checkmark & \checkmark &  &  & 87.42 $\pm$ 0.59 & 78.42 $\pm$ 0.87 & 95.56 $\pm$ 0.73 \\
  \checkmark &  & \checkmark &  & 87.33 $\pm$ 0.48 & 78.34 $\pm$ 0.70 & 95.44 $\pm$ 0.53 \\
  \checkmark & \checkmark & \checkmark &  & 87.62 $\pm$ 0.54& 78.75 $\pm$ 0.82 & \textbf{95.84 $\pm$ 0.61} \\
  \checkmark &  &  & \checkmark & 87.09 $\pm$ 0.59 & 77.97 $\pm$ 0.93 & 95.34 $\pm$ 0.69 \\
  \checkmark & \checkmark & \checkmark & \checkmark & \textbf{87.65 $\pm$ 0.57}& \textbf{78.78 $\pm$ 0.83} & 95.62 $\pm$ 0.59 \\
  \hline
  \hline
  \end{tabular} 
  \label{ablation}
  \vspace{-0.0cm}
\end{table}

The performance of our UGMCS-Net and its four variants are listed in Table~\ref{ablation}, `-' indicates Attention U-Net.
V1 refers to UGS-Net with no IUCM and Multiple Annotation Fusion Loss, which serves as the base network for the other variants.
The results show that:
(1) The performance improvements of UGMCS-$\Phi_a$ and UGMCS-$\Phi_b$ networks over UGMCS-Net indicate that the fusion of all annotated information can enable the network to more accurately capture the nodule region and achieve better segmentation performance.
(2) The DSC, IoU, and NSD of UGMCS-$\Phi_a$+$\Phi_b$ network are higher than those of the UGMCS-$\Phi_a$ and UGMCS-$\Phi_b$ networks, indicating that using Multiple Annotation Fusion Loss for both UAM and the final output simultaneously is better than using either one alone.
(3) Our UGMCS-Net outperforms UGMCS-$\Phi_a$+$\Phi_b$ and UGMCS-IUCM, demonstrating the effectiveness of the Intersection-Union Constraining Module. Although the quantitative performance of the network slightly declined when only the Intersection-Union Constraining Module was added (UGMCS-IUCM), a significant qualitative improvement was observed, which will be discussed later.
Moreover, the superior performance of UGMCS-Net suggests that the Multiple Annotation Fusion Loss and Intersection-Union Constraining Module can mutually enhance each other, constraining uncertainty and facilitating better segmentation performance. 

\begin{figure*}[h]
  \centerline{\includegraphics[width=160mm]{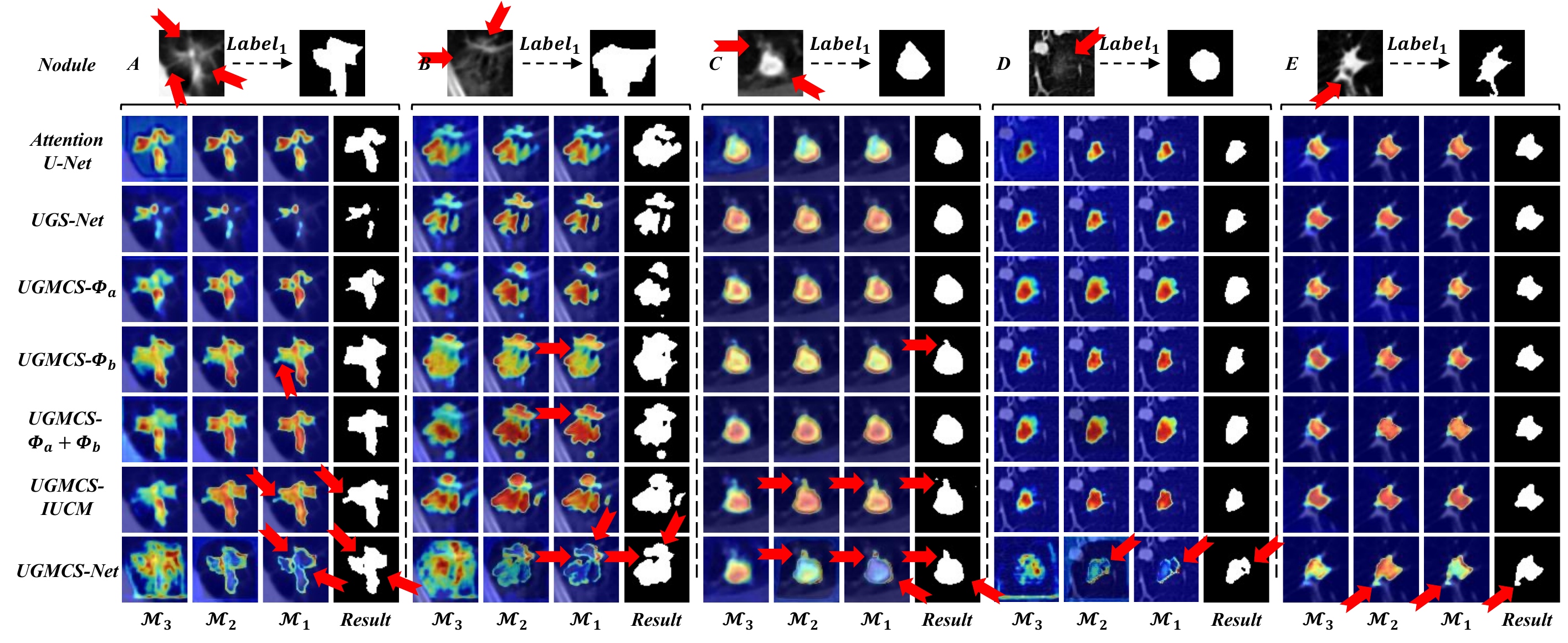}}
  \vspace{-0.0cm}
  \caption{Visualization of different convolutional layers.
  The \textbf{\emph{Result}} in each case represents the final prediction of the network.
  \textbf{\emph{M3}}, \textbf{\emph{M2}}, and \textbf{\emph{M1}} respectively represent the visual feature maps of the third from the bottom, the second, and the first convolution layer under different network configurations.
  Red arrows indicate areas of nodules that need attention.}
  \vspace{-0.0cm}
  \label{F2}
\end{figure*}

\begin{figure}[h]
  \centerline{\includegraphics[width=85mm]{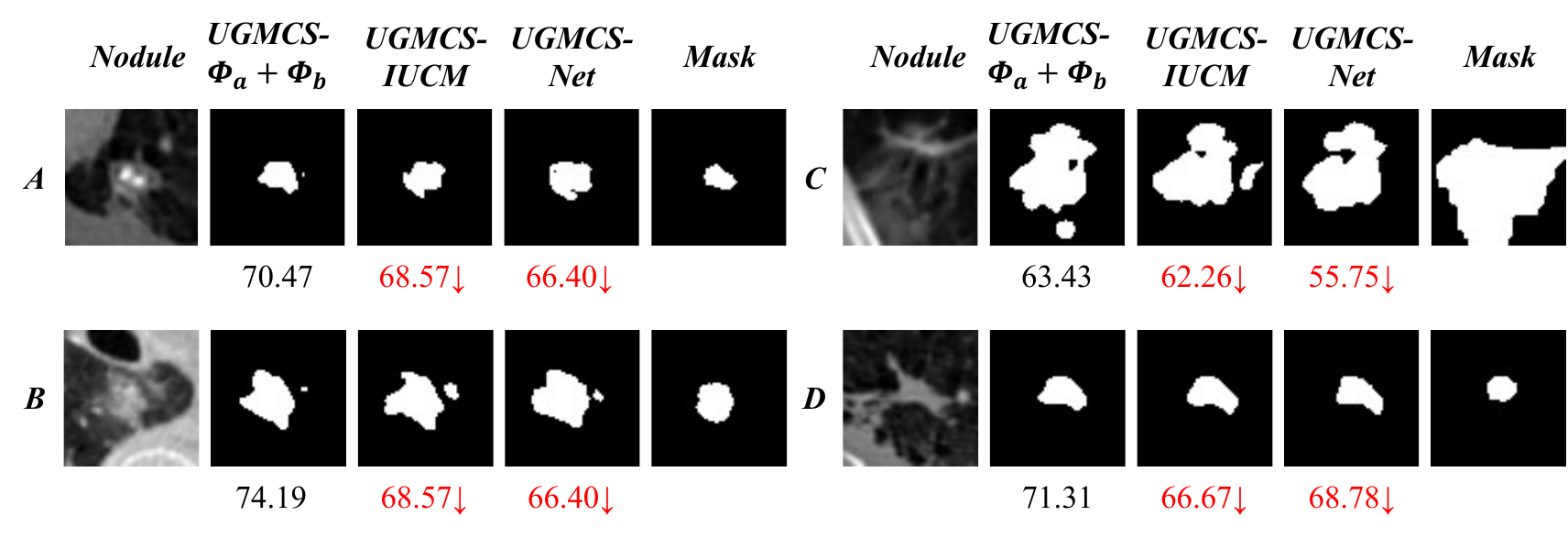}}
  \vspace{-0.0cm}
  \caption{Failure Cases Demonstration. The number below each segmentation mask is its DSC score. The unit of DSC is the percentage.}
  \vspace{-0.1cm}
  \label{F3}
\end{figure}

To further verify the validity of the Multiple Annotation Fusion Loss and Intersection-Union Constraining Module, we demonstrate feature map visualization using Grad-CAM \cite{selvaraju2017grad} in Figure~\ref{F2}.
The \textbf{\emph{Result}} in each case represents the final prediction of the network.
\textbf{\emph{M3}}, \textbf{\emph{M2}}, and \textbf{\emph{M1}} respectively represent the visual feature maps of the third from the bottom, the second and the first convolution layer under different network configurations.
Based on Figure~\ref{F2}, we observe that:
(1) When Multiple Annotation Fusion Loss is applied to the $\cup(X)$ or $X_S$, the recognition ability of the network for the low-density organization was significantly improved (UGMCS-$\Phi_a$ network and UGMCS-$\Phi_b$ network, Nodule A-D);
(2) The simultaneous use of Multiple Annotation Fusion Loss in $\cup(X)$ and $X_S$ can make the network outline the nodule boundary more clearly based on improving the sensitivity to low-density tissues (UGMCS-$\Phi_a$+$\Phi_b$ network, Nodule A-D).
(3) Intersection-Union Constraining Module enables the network to learn more boundary features, such as the spiculation (UGMCS-IUCM network, Nodule A-C).   
(4) When Multiple Annotation Fusion Loss and Intersection-Union Constraining Module are used at the same time, the network's attention shifts to the nodule boundary, delineating a more reasonable and complete nodule region (UGMCS-Net, Nodule A-E).

As shown in Figure~\ref{F2}, with the help of IUCM, the network can obtain more semantic and reasonable segmentation results for complex nodules.
However, as shown in Table~\ref{ablation}, UGMCS-Net has weak performance gains on DSC, IoU, and NSD compared to UGMCS-$\Phi_a$+$\Phi_b$.
We believe there are three reasons for this phenomenon:
(1) IUCM focuses on improving the segmentation performance of complex nodules, and compared with the improvement in measurement, IUCM makes the model improve the performance of segmentation results more significantly (further verification is provided in Section\ref{CNV}).
(2) Complex nodules only account for a small part of the total number of nodules, so IUCM cannot significantly improve the model's score on various metrics.
(3) There are still some failure cases in the dataset. As shown in Figure~\ref{F3}, in these cases, the segmentation masks obtained by UGMCS-IUCM and UGMCS-Net contained more nodular tissue and were more accurate lesion areas, but their DSC scores were lower.
Since the knowledge bias of different doctors' fields will affect the ground truth, obtaining a more accurate and reasonable segmentation mask is often more meaningful than a higher metric score for segmentation tasks.

\subsubsection*{Ablation Study for Backbone}
We test U-Net and R2U-Net as the backbone of the three modules.
As listed in Table~\ref{backbone_ablation}, the model with U-Net as the backbone obtained 87.04$\%$, 78.07$\%$, and 94.50$\%$ scores on DSC, IoU, and NSD, respectively.
The model with R2U-Net as the backbone obtained 86.20$\%$, 76.82$\%$, and 93.75$\%$ results on DSC, IoU, and NSD, respectively.
Experimental results demonstrate that the proposed Feature Extracting Module, Uncertainty-Aware Module, and Intersection-Union Constraining Module are plug-and-play. 
Our previous work evaluated nnU-Net [22] as the backbone. This network has two drawbacks in this task: (1) it took too much calculation resources, and (2) we only have 1860 nodules, which may lead to over-fitting. To balance the calculation resources and performance gains, we choose Attention U-Net in this work.

\begin{table}[htbp]
  \setlength{\tabcolsep}{10pt}
  \renewcommand\arraystretch{1.1}
  \vspace{-0.0cm}
  \caption{Performance comparison between different backbone.}
    \centering
  \begin{tabular}{c|c|c|c}		
  \hline
  \hline
  Model & DSC & IoU & NSD \\
  \hline
  U-Net & 86.26$\%$ & 76.79$\%$ & 94.46$\%$\\
  Upgraded U-Net & 87.04$\%$& 78.07$\%$ &94.50$\%$\\
  R2U-Net & 85.17$\%$ & 75.29$\%$ & 94.03$\%$\\
  Upgraded R2U-Net& 86.20$\%$ & 76.82$\%$ & 93.75$\%$\\
  Attention U-Net& 86.67$\%$ & 77.33$\%$ & 94.94$\%$\\
  UGMCS-Net & 87.65$\%$ & 78.78$\%$ & 95.62$\%$\\
  \hline
  \hline
  \end{tabular} 
  \label{backbone_ablation}
  \vspace{-0.0cm}
\end{table}

\subsubsection*{Ablation Study for Filters in Intersection-Union Constraining Module}
In Section~\ref{SCM}, we discussed the sensitivity of Otsu to density characteristics, enabling the model to identify high-density tissue accurately. Consequently, we utilize Otsu's method for the feature extraction of $R_{HC}$.
Conversely, Gabor, being sensitive to image edges and offering effective direction and scale selection features, is selected for the feature extraction of $R_{LC}$.
Within these sections, we conducted experiments involving five filter settings within the Intersection-Union Constraining Module on Fold1. These settings are outlined in Table~\ref{filters_choice}.
As demonstrated in Table~\ref{filters_choice}, our final configuration yields the highest DSC score.

\begin{table}[h]
  \setlength{\tabcolsep}{10pt}
  \renewcommand\arraystretch{1.1}
  \vspace{-0.0cm}
  \caption{Performance comparison between different filter settings.}
    \centering
  \begin{tabular}{c|c|c|c|c}		
  \hline
  \hline
  Index & $R_{HC}$& $R_{LC}$ & $R_{Uni}$ & DSC\\
  \hline
  setting1 & Otsu& Otsu &Otsu& 87.06$\%$\\
  setting2 & Gabor& Gabor &Gabor& 87.30$\%$\\
  setting3 & Gabor& Otsu &Gabor& 87.37$\%$\\
  setting4 & Gabor&Gabor&Otsu& 87.28$\%$\\
  \textbf{final setting} & \textbf{Otsu} & \textbf{Gabor} & \textbf{Gabor} & \textbf{87.49$\%$}\\
  \hline
  \hline
  \end{tabular} 
  \label{filters_choice}
  \vspace{-0.0cm}
\end{table}

\subsubsection*{Ablation Study for Parameters}
In Equation~\ref{eq_weight}, three manually set parameters are present: $\alpha_1$ is designated as 0.5, $\alpha_2$ as 0.5, and $\alpha_3$ as 1. In this section, we undertake a five-fold validation and illustrate the rationale behind these parameter selections.
As depicted in Table~\ref{three_alpha}, the proposed method attains its peak performance when $\alpha_1$ is set to 0.5, $\alpha_2$ to 0.5, and $\alpha_3$ to 1. Notably, $\alpha_3$ represents the weight attributed to the final segmentation, which means $X_S$ plays an important role during training.

Moreover, we perform experiments about the probability map weighted average fusion for the three branches, hoping the network can choose proper weights for each branch in IUCM. However, we observe that the branch corresponding to $R_{Uni}$ could provide more generalized features for nodule segmentation and appeared 'stronger' than the other two branches. Consequently, the weights for the other two branches tend to reduce to 0. Due to these observations, we choose to fuse the probability maps directly.

\begin{table}[h]
  \setlength{\tabcolsep}{11pt}
  \renewcommand\arraystretch{1.1}
  \vspace{-0.0cm}
  \caption{Performance comparison between different $\alpha$ settings.}
    \centering
  \begin{tabular}{c|c|c|c|c|c}		
  \hline
  \hline
  \multicolumn{3}{c}{\textbf{Parameters}} & \multicolumn{3}{c}{\textbf{Performances}}\\
  \hline
  $\alpha_1$&$\alpha_2$&$\alpha_3$& DSC & IoU & NSD\\
  \hline
  1& 1 &1& 86.78$\%$ & 77.69$\%$ & 94.61$\%$\\
  1& 0.5 &0.5& 87.23$\%$ & 78.34$\%$ & 94.44$\%$\\
  0.5& 1 &0.5& 87.10$\%$ & 78.02$\%$ & 94.58$\%$\\
  0.5& 0.5 &0.5& 86.99$\%$ & 77.96$\%$ & 93.56$\%$\\
  \textbf{0.5} & \textbf{0.5} & \textbf{1} & \textbf{87.65$\%$} & \textbf{78.78$\%$} & \textbf{95.62$\%$} \\
  \hline
  \hline
  \end{tabular} 
  \label{three_alpha}
  \vspace{-0.0cm}
\end{table}

\begin{table*}[h]
  \renewcommand\arraystretch{1.2}
  \vspace{-0.0cm}
  \caption{Analysis of segmentation performances in Complex Nodule Validation.
  UGMCS-$\Phi_a$+$\Phi_b$, UGMCS-IUCM, and UGMCS-Net average DSC and IoU are followed by the difference (green number) from the corresponding metric of Attention U-Net.
  All indicators are expressed in percentages.}
  \centering
  \begin{tabular}{c|cc|cc|cc|cc|cc|cc}
  \hline  
  \hline
  & \multicolumn{2}{c}{\textbf{\emph{Fold1}}} & \multicolumn{2}{c}{\textbf{\emph{Fold2}}} & \multicolumn{2}{c}{\textbf{\emph{Fold3}}} & \multicolumn{2}{c}{\textbf{\emph{Fold4}}} & \multicolumn{2}{c}{\textbf{\emph{Fold5}}} & \multicolumn{2}{c}{\textbf{\emph{Average}}}\\
  \hline
  & \textbf{\emph{DSC}} & \textbf{\emph{IoU}} & \textbf{\emph{DSC}} & \textbf{\emph{IoU}} & \textbf{\emph{DSC}} & \textbf{\emph{IoU}} & \textbf{\emph{DSC}} & \textbf{\emph{IoU}} & \textbf{\emph{DSC}} & \textbf{\emph{IoU}} & \textbf{\emph{DSC}} & \textbf{\emph{IoU}} \\
  \hline
  &\multicolumn{12}{c}{\textbf{\emph{U-Net DSC $\leq$ 60$\%$}}} \\
  \hline
  \makecell[l]{\textbf{\emph{U-Net}}} & 46.24 & 30.55 & 51.51 & 35.00 & 52.18 & 35.35 & 46.11 & 30.84 & 46.6 & 30.96 & 48.53 & 32.36 \\
  \makecell[l]{\textbf{\emph{Attention U-Net}}} & 64.30 & 48.08 & 58.07 & 42.36 & 53.13 & 36.28 & 53.28 & 40.66 & 41.74 & 26.78 & 54.10 & 38.83 \\
  \makecell[l]{\textbf{\emph{UGMCS-$\Phi_a$+$\Phi_b$}}} & 55.10 & 39.11 & 47.60 & 33.82 & 61.54 & 44.93 & 63.27 & 48.68 & 50.88 & 34.27 & 55.68\textcolor{green}{$\uparrow1.58$} & 40.16\textcolor{green}{$\uparrow1.33$} \\
  \makecell[l]{\textbf{\emph{UGMCS-IUCM}}} & 66.35 & 51.08 & 55.38 & 40.27 & 65.89 & 52.17 & 57.33 & 42.52 & 51.14 & 34.46 & 59.22\textcolor{green}{$\uparrow5.12$} & 44.10\textcolor{green}{$\uparrow5.27$} \\
  \makecell[l]{\textbf{\emph{UGMCS-Net}}} & 70.42 & 55.5 & 56.95 & 41.57 & 63.17 & 47.68 & 46.64 & 32.52 & 60.59 & 44.94 & 59.55\textcolor{green}{$\uparrow5.45$} & 44.44\textcolor{green}{$\uparrow5.61$} \\
  \hline
  &\multicolumn{12}{c}{\textbf{\emph{U-Net DSC $\leq$ 70$\%$}}} \\
  \hline
  \makecell[l]{\textbf{\emph{U-Net}}} & 56.88 & 40.78 & 61.37 & 44.79 & 61.54 & 44.83 & 61.08 & 44.78 & 59.53 & 43.24 & 60.08 & 43.68 \\
  \makecell[l]{\textbf{\emph{Attention U-Net}}} & 69.44 & 54.26 & 63.50 & 47.93 & 64.75 & 48.64 & 66.27 & 52.33 & 62.64 & 47.28 & 65.32 & 50.09 \\
  \makecell[l]{\textbf{\emph{UGMCS-$\Phi_a$+$\Phi_b$}}} & 61.50 & 45.87 & 61.98 & 47.82 & 69.47 & 54.61 & 70.22 & 55.59 & 66.88 & 51.68 & 66.01\textcolor{green}{$\uparrow0.69$} & 51.11\textcolor{green}{$\uparrow1.02$} \\
  \makecell[l]{\textbf{\emph{UGMCS-IUCM}}} & 70.14 & 55.76 & 66.00 & 51.20 & 69.60 & 54.71 & 69.11 & 55.11 & 67.29 & 52.11 & 68.43\textcolor{green}{$\uparrow3.11$} & 53.78\textcolor{green}{$\uparrow3.69$} \\
  \makecell[l]{\textbf{\emph{UGMCS-Net}}} & 74.09 & 60.02 & 69.32 & 54.74 & 72.02 & 57.62 & 65.90 & 50.99 & 70.60 & 56.05 & 70.39\textcolor{green}{$\uparrow5.07$} & 55.88\textcolor{green}{$\uparrow5.79$} \\
  \hline
  &\multicolumn{12}{c}{\textbf{\emph{U-Net DSC $\leq$ 80$\%$}}} \\
  \hline
  \makecell[l]{\textbf{\emph{U-Net}}}& 70.04 & 55.00 & 72.58 & 57.52 & 72.60 & 57.50 & 70.71 & 55.46 & 71.42 & 56.30 & 71.47 & 56.36 \\
  \makecell[l]{\textbf{\emph{Attention U-Net}}} & 75.55 & 61.47 & 74.11 & 59.89 & 76.45 & 62.92 & 76.02 & 62.73 & 75.06 & 61.01 & 75.44 & 61.60 \\
  \makecell[l]{\textbf{\emph{UGMCS-$\Phi_a$+$\Phi_b$}}} & 73.79 & 59.93 & 74.24 & 60.85 & 77.00 & 63.52 & 75.57 & 61.77 & 77.29 & 63.97 & 75.58\textcolor{green}{$\uparrow0.14$} & 62.01\textcolor{green}{$\uparrow0.41$} \\
  \makecell[l]{\textbf{\emph{UGMCS-IUCM}}} & 76.34 & 62.85 & 75.77 & 62.13 & 78.13 & 65.01 & 76.55 & 63.22 & 78.02 & 64.97 & 76.96\textcolor{green}{$\uparrow1.52$} & 63.64\textcolor{green}{$\uparrow2.04$} \\
  \makecell[l]{\textbf{\emph{UGMCS-Net}}} & 79.11 & 66.21 & 77.16 & 63.63 & 77.27 & 63.70 & 74.49 & 60.64 & 77.53 & 64.00 & 77.08\textcolor{green}{$\uparrow1.64$} & 63.64\textcolor{green}{$\uparrow2.04$} \\
  \hline
  \hline
  \end{tabular} 
  \label{complexNoduleTable}
  \vspace{-0.0cm}
\end{table*}

\begin{figure*}[htbp]
  \centerline{\includegraphics[width=160mm]{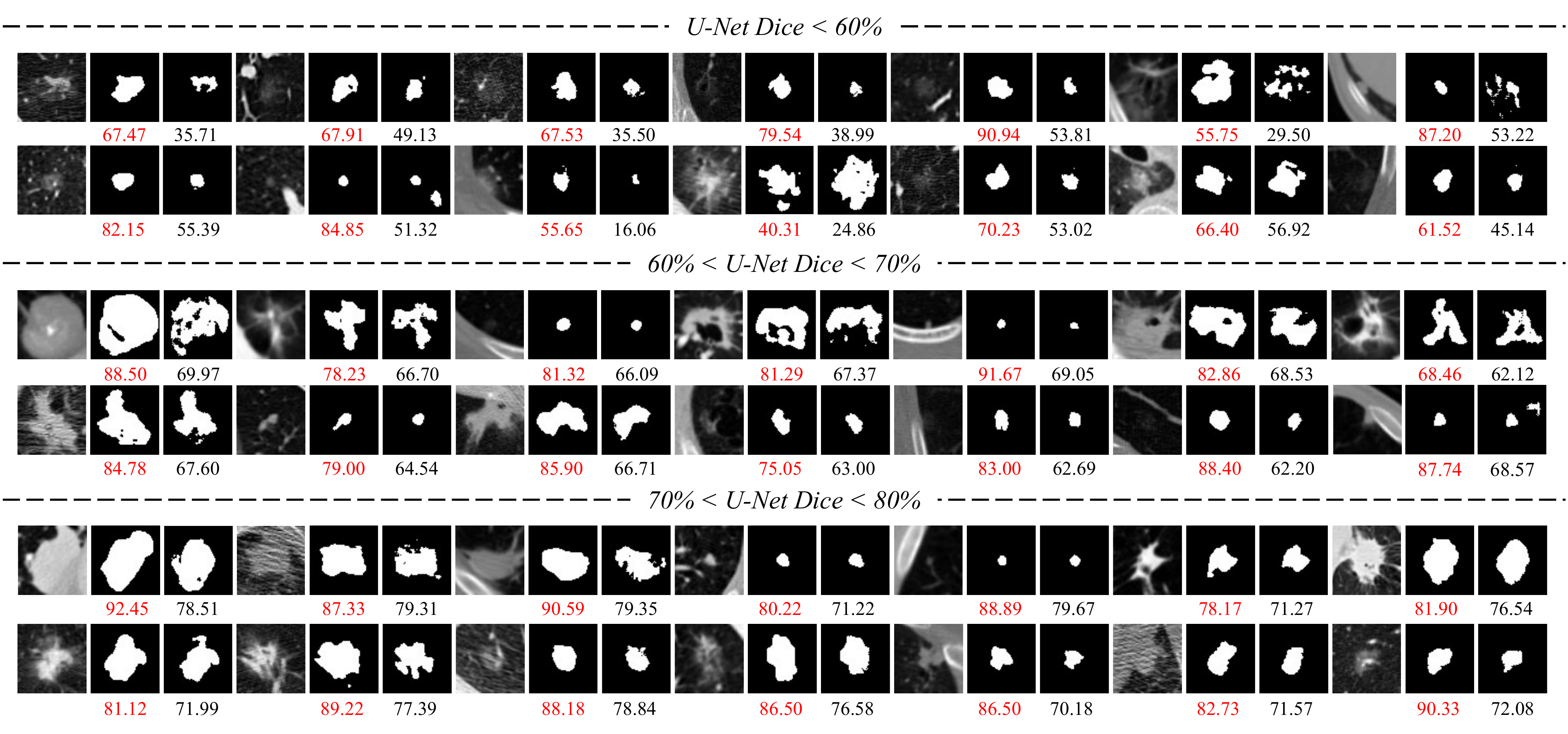}}
  \vspace{-0.0cm}
  \caption{Complex-Nodule Validation.
  This validation tests UGMCS-Net's segmentation performance on the lung nodules that are difficult to segment by U-Net by three levels.
  The last two masks of each CT image are the segmentation results of UGMCS-Net and U-Net respectively.
  The segmentation results of U-Net are shown in black, and the UGMCS-Net are shown in red.
  All indicators are expressed in percentages.}
  \vspace{-0.0cm}
  \label{complexNodule}
\end{figure*}

\subsection{Complex-Nodule Validation}
\label{CNV}

The UGMCS-Net learns features from regions that may cause segmentation uncertainty. As a result, it can better segment nodules with large low-dense regions or complex structures. To better demonstrate its improvement on nodules that are difficult to segment by U-Net, we design a Complex-Nodule Validation.
Based on the five-fold verification U-Net, we further selected three groups of nodules with DSC scores below 60$\%$, 70$\%$, and 80$\%$. The Attention U-Net, UGMCS-$\Phi_a$+$\Phi_b$, UGMCS-IUCM, and UGMCS-Net are then trained with the same data settings to test these nodules again and compare the DSC and IoU scores.

The improvement in the performance of UGMCS-Net on complex nodules is evident. In comparison to Attention U-Net, for nodules with DSC scores below 60$\%$ on U-Net, UGMCS-$\Phi_a$+$\Phi_b$ leads to an average DSC score increase of 1.58$\%$, UGMCS-IUCM elevates the average DSC score by 5.12$\%$, and UGMCS-Net yield an average DSC score increase of 5.45$\%$.
For nodules with DSC scores ranging between 60$\%$ and 70$\%$ on U-Net, UGMCS-$\Phi_a$+$\Phi_b$, UGMCS-IUCM, and UGMCS-Net achieved average DSC score improvements of 0.69$\%$, 3.11$\%$, and 5.07$\%$, respectively.
Similarly, for nodules with DSC scores between 70$\%$ and 80$\%$ on U-Net, UGMCS-$\Phi_a$+$\Phi_b$, UGMCS-IUCM, and UGMCS-Net exhibited average DSC score enhancements of 0.14$\%$, 1.52$\%$, and 1.64$\%$, correspondingly.
It can be seen from the above difference that UGMCS-IUCM has a great degree of performance improvement for complex nodules.
According to the table \ref{tabel1}, our network improves DSC by only 0.89$\%$ over the entire dataset, but has a large performance boost relative to Attention U-Net on nodules with DSC scores less than 60 $\%$ on U-Net.
This is because nodules with complex structures make up only a small fraction of all data.

Figure \ref{complexNodule} shows the segmentation results of some complex nodules. 
The red subscripts are the segmentation DSC of UGMCS-Net, and the black subscripts are the DSC of U-Net. 
It can be perceived that nodules with U-Net segmentation DSC score less than 60$\%$ are some low-density or ground glass tissues. The significant improvements of UGMCS-Net in these nodules indicate that UGMCS-Net can better learn the characteristics of nodules' low-density tissue and more accurately segment low-density nodular lesion regions.
When the U-Net segmentation DSC score is less than 70$\%$, it can be observed that in addition to some low-density nodules, some nodules also have irregular cavities, spiculation, abrupt bright spots in the tissues, or over-bright lung walls. The convincing segmentation performance of UGMCS-Net on these nodules reflects the learning ability of UGMCS-Net on boundary features, density differences, and good anti-noise ability.
When the U-Net segmentation DSC score is less than 80$\%$, we observe that many new nodules that have more solid tissues. In this case, UGMCS-Net can accurately determine the nodule region. In addition, for most nodules, the segmentation results of UGMCS-Net reflect more semantic features and have stronger interpretability.

\section{Discussions}
\label{discussions}
The task of lung nodule segmentation has focused on achieving high accuracy for a very long time, with the DSC or IoU being the main target. However, given the small size and complex structures of lung nodules, it may be more helpful for radiologists if segmentation results highlight regions with varying degrees of confidence. High-confidence regions provide the major part of nodules or tumor tissues, while low-confidence regions contain important low-density features, such as ground-glass and spiculation signs, that radiologists should also pay attention to.

The proposed method aims to provide information that is more useful for clinical diagnosis, rather than simply improving the DSC. It does not seek to replace the clinical role of doctors, but to complement it by allowing them to leverage the strengths of artificial intelligence methods. This, we believe, is a better approach to integrating AI into clinical practice.

\subsection{Data Requirement}
Our approach does not necessitate every radiologist to annotate all nodules. As indicated in \cite{armato2011lung}, a collective total of 12 radiologists took part in the image annotation procedure across all five sites during the project. Given that most nodules are annotated by 1-4 radiologists, it is conceivable that nodules might be annotated by various radiologists.

Nonetheless, the involvement of different radiologists in annotating nodules will not impede the applicability of our method. Despite potential variance in annotation style among different radiologists, it's worth noting that well-trained radiologists follow the established standards for nodule annotation such as \cite{bankier2017recommendations}. These standards ensure that different groups of radiologists offer diverse but largely consistent annotations.

\subsection{Limitation}
One limitation of our study is that we only test our proposed method on the LIDC-IDRI dataset, currently the only publicly available fully annotated dataset for lung nodules. Despite being over ten years old, this dataset still presents many opportunities for research. However, the need for multiple annotations could limit the practicality of our method in real-world clinical settings, where obtaining multiple annotations may not always be feasible. To address this limitation, we plan to explore techniques that can automatically identify high- and low-confidence regions based on a single annotation, thus increasing the feasibility and applicability of our approach.

\section{Conclusions}
\label{conclusion}
This paper introduces the Uncertainty-Aware Attention Mechanism (UAAM), which leverages the consensus or disagreements among multiple annotations to improve segmentation and identify regions with low segmentation confidence. UAAM captures features from the Multi-Confidence Mask (MCM), a combination of a Low-Confidence (LC) Mask and a High-Confidence (HC) Mask. 
Based on UAAM, we further design an Uncertainty-Guide Segmentation Network (UGMCS-Net), which contains a \emph{Feature Extracting Module}, an \emph{Uncertainty-Aware Module}, and an \emph{Intersection-Union Constraining Module}. These modules together learn valuable information from the consensus or disagreements among multiple annotations, providing regions with high and low segmentation confidences, and a segmentation result that can balance all possibilities.
Besides the traditional validation method, we propose a Complex Nodule Validation on LIDC-IDRI, which tests UGMCS-Net's segmentation performance on the lung nodules that are difficult to segment by U-Net. Experimental results demonstrate that our method can significantly improve the segmentation performance on nodules with poor segmentation by U-Net.

\ifCLASSOPTIONcaptionsoff
  \newpage
\fi



%
  \bibliographystyle{IEEEtran}
  \bibliography{ijcai20}

%






\end{document}